\documentclass[prd,twocolumn,showpacs,superscriptaddress]{revtex4-1}
\usepackage{graphicx}
\usepackage{units}
\usepackage{multirow}
\usepackage{bigstrut}
\usepackage{overpic}
\usepackage{array}
\usepackage{color}
\usepackage{amsfonts}
\usepackage{amsmath}
\usepackage{amssymb}
\usepackage{amsthm}
\usepackage{bm}
\usepackage{times}
\usepackage{lipsum}
\usepackage{bigints}
\usepackage{dcolumn}
\usepackage{ulem}
\usepackage{bm}
\usepackage[colorlinks,citecolor=blue,urlcolor=blue,linkcolor=blue]{hyperref}
\begin{document}

\setlength{\oddsidemargin}{-0.5cm} \addtolength{\topmargin}{15mm}

\title{\boldmath Improved model-independent determination of the strong-phase difference between $D^{0}$ and $\bar{D}^{0}\to K^{0}_{\mathrm{S,L}}K^{+}K^{-}$ decays}

\author{
\begin{small}
\begin{center}
M.~Ablikim$^{1}$, M.~N.~Achasov$^{10,c}$, P.~Adlarson$^{64}$, S. ~Ahmed$^{15}$, M.~Albrecht$^{4}$, R.~Aliberti$^{28}$, A.~Amoroso$^{63A,63C}$, Q.~An$^{60,48}$, ~Anita$^{21}$, X.~H.~Bai$^{54}$
% ...
\\
\vspace{0.2cm}
(BESIII Collaboration)\\
\vspace{0.2cm} {\it
$^{1}$ Institute of High Energy Physics, Beijing 100049, People's Republic of China\\
% ...
$^{4}$ Bochum Ruhr-University, D-44780 Bochum, Germany\\
% ...
$^{10}$ G.I. Budker Institute of Nuclear Physics SB RAS (BINP), Novosibirsk 630090, Russia\\
% ...
$^{15}$ Helmholtz Institute Mainz, Johann-Joachim-Becher-Weg 45, D-55099 Mainz, Germany\\
% ...
$^{21}$ Indian Institute of Technology Madras, Chennai 600036, India\\
% ...
$^{28}$ Johannes Gutenberg University of Mainz, Johann-Joachim-Becher-Weg 45, D-55099 Mainz, Germany\\
% ...
$^{48}$ State Key Laboratory of Particle Detection and Electronics, Beijing 100049, Hefei 230026, People's Republic of China\\
% ...
$^{54}$ University of Jinan, Jinan 250022, People's Republic of China\\
% ...
$^{60}$ University of Science and Technology of China, Hefei 230026, People's Republic of China\\
% ...
$^{63}$ (A)University of Turin, I-10125, Turin, Italy; (B)University of Eastern Piedmont, I-15121, Alessandria, Italy; (C)INFN, I-10125, Turin, Italy\\
$^{64}$ Uppsala University, Box 516, SE-75120 Uppsala, Sweden\\
% ...
$^{c}$ Also at the Novosibirsk State University, Novosibirsk, 630090, Russia\\
% ...
}
\end{center}
\vspace{0.4cm}
\end{small}
}

\begin{abstract}
We present a measurement of the strong-phase difference between $D^{0}$ and $\bar{D}^{0}\to K^{0}_{\rm S,L}K^{+}K^{-}$ decays, performed through a study of quantum-entangled pairs of charm mesons. The measurement exploits a data sample equivalent to an integrated luminosity of 2.93~fb$^{-1}$, collected by the BESIII detector in $e^{+}e^{-}$ collisions corresponding to the mass of the  $\psi(3770)$ resonance.  The strong-phase difference is an essential input to the determination of the Cabibbo-Kobayashi-Maskawa (CKM) angle $\gamma/\phi_3$ through the decay $B^{-}\to DK^{-}$, where $D$ can be either a $D^{0}$ or a $\bar{D}^{0}$ decaying to $K^{0}_{\rm S,L}K^{+}K^{-}$. This is the most precise measurement to date of the strong-phase difference in these decays.

\end{abstract}


\maketitle

\section{\boldmath Introduction}\label{sec:introduction}
In the Standard Model (SM), the charged-weak interaction in the quark sector is described by the Cabibbo-Kobayashi-Maskawa matrix (CKM)~\cite{ckmmatrix}. One of the primary goals of flavor physics experiments is to determine the angles $\alpha, \beta$~and~$\gamma$ (or  $\phi_{2},\phi_{1}$ and $\phi_{3}$) of the $b-d$ CKM unitary triangle precisely. Currently, the most precise measurements of $\gamma$ are extracted using tree-level $B^{-}\to DK^{-}$ decays~\cite{GLW}. Here and elsewhere in this paper $D$ refers to either a $D^{0}$ or a $\bar{D}^{0}$ meson decaying into the same final state and charge conjugation is implicit, unless stated otherwise. 
The sensitivity to $\gamma$ arises from the interference of two amplitudes: $b\to c\bar{u}s$ that results in the $B^{-}\to D^{0}K^{-}$ decay, and $b\to u\overline{c}s$ that leads to the $B^{-}\to \bar{D}^{0}K^{-}$ decay. The latter amplitude is both CKM- and color-suppressed relative to the former.
The value of $\gamma$ measured with such tree-level transitions is insensitive to loop-level contributions \cite{bib:brod}. Therefore, tests for new physics that are made by comparing unitarity triangle parameters measured using tree and loop processes can be improved by more precise determinations of $\gamma$ \cite{bib:buras,bib:lenz}. 

Different methods of determining $\gamma$ are classified based upon the decay products of the $D$ decay: $CP$ eigenstates (GLW method)~\cite{GLW}, flavor-eigenstates (ADS method)~\cite{ADS}, and self-conjugate multibody states (BPGGSZ method)~\cite{bondarK0Spipi,giri,bellegamma}. The most widely used $D$ decays for the BPGGSZ method are $D\to K^{0}_{\rm{S}}h^{+}h^{-}$, where $h = \pi, K$. Measurements of $\gamma$ using these final states have been performed by the Belle, BaBar and LHCb Collaborations~\cite{bellegamma,babargamma,LHCbgamma}. Recently the first constraints on $\gamma$ using the BPGGSZ method with a four-body $D$ decay were reported~\cite{resmi}. BPGGSZ analyses require an understanding of the interference effects between $D^{0}$ and $\bar{D}^{0}$ decays, especially concerning the strong-phase difference between the $D^{0}$ and $\bar{D}^{0}$ decay amplitudes.

A precise measurement of the strong-phase difference in $D\to K^{0}_{\rm S, L}\pi^{+}\pi^{-}$ decays was reported by the BESIII Collaboration recently~\cite{lei}. The first measurements of the strong-phase difference between $D^{0}$ and $\bar{D}^{0}$ decaying to the $K^{0}_{\rm S,L}K^{+}K^{-}$ final state were reported by the CLEO Collaboration, using a data set equivalent to an integrated luminosity of 818 pb$^{-1}$ that was collected at a center-of-mass energy corresponding to the mass of the $\psi(3770)$ resonance~\cite{jim}. In this paper, we present an improved measurement of the strong-phase parameters for $D\to K^{0}_{\rm S, L}K^{+}K^{-}$ decays, using a $\psi(3770)$ data sample corresponding to an integrated luminosity of 2.93 fb$^{-1}$ recorded by the BESIII detector. This measurement can be used as an input to the model-independent measurement of $\gamma$ using the BPGGSZ method. Moreover, these strong-phase parameters serve as an essential input to the model-independent determination of charm-mixing parameters and in probing $CP$ violation with $D\to K^{0}_{\rm S, L}K^{+}K^{-}$ decays~\cite{guy}. 

The $D \to K^{0}_{\rm S}K^{+}K^{-}$ decay proceeds via various intermediate resonances, which leads to a significant strong-phase variation over the phase space. We define the kinematic variables $m_{\pm}^{2} = (P_{K_{\rm S}^{0}} + P_{K_{\pm}})^{2}$, which serve as the basis of the $D \to K^{0}_{\rm S}K^{+}K^{-}$ Dalitz plot. Here, $P_i~(i = K_{\rm S}^{0},K^{+},K^{-})$ is the four-momentum of the $D$ decay product. The amplitude for $B^{-} \to D(K^{0}_{\rm S}K^{+}K^{-}) K^{-}$ at $(m^{2}_{+},m^{2}_{-})$ can be written as
\begin{eqnarray}
f_{B^{-}}\left(m^{2}_{+},m^{2}_{-}\right) & \propto &  f_D\left(m^{2}_{+},m^{2}_{-}\right) \nonumber\\ && + r_B e^{i\left(\delta_B-\gamma\right)} f_{\bar{D}}(m^{2}_{+},m^{2}_{-})\;, 
\label{eq:ampbdec1}
\end{eqnarray}
 where $r_{B}$ is the ratio of the magnitude of the suppressed to the favored $B$-decay amplitude,  $\delta_B$ is the $CP$-conserving strong-phase difference between favored and suppressed $B$-decay amplitudes, $\gamma$ is the weak-phase difference between the $B$ decay amplitudes, and $f_{D}(m^{2}_{+},m^{2}_{-})$ $\left(f_{\bar{D}}(m^{2}_{+},m^{2}_{-})\right)$ is the amplitude of the $D^{0}\to~K^{0}_{\rm S}K^{+}K^{-}~(\bar{D}^{0}\to~K^{0}_{\rm S}K^{+}K^{-})$ decay. We neglect $CP$ violation in $D$ decays as in Ref.~\cite{giri}, and can thus use the relation $f_{\bar{D}}(m^{2}_{+},m^{2}_{-}) = f_{D}(m^{2}_{-},m^{2}_{+})$ so that  Eq.~(\ref{eq:ampbdec1}) can be written as
\begin{equation}
f_{B^{-}}(m^{2}_{+},m^{2}_{-})\propto f_D(m^{2}_{+},m^{2}_{-}) + r_B e^{i(\delta_B-\gamma)} f_{D}(m^{2}_{-},m^{2}_{+}).
\label{eq:ampbdec2}
\end{equation}
Therefore, the decay rate of a $B^{-}$ meson is
\begin{widetext}
\begin{equation}
\Gamma_{{B}^{-}}\left(m_{+}^{2},m_{-}^{2}\right)\propto \left|f_{D}\left(m_{+}^{2},m_{-}^{2}\right)\right|^{2} + r_{B}^{2}\left|f_{D}\left(m_{-}^{2},m_{+}^{2}\right)\right|^{2} + 2r_{B} \left|f_{D}(m_{+}^{2},m_{-}^{2})\right|\left|f_{D}(m_{-}^{2},m_{+}^{2})\right|\cos{\left(\Delta\delta_{D} + \delta_{B} - \gamma\right)} \;,
 \label{eq:bdecay_rate}
\end{equation}
\end{widetext}
where $\Delta\delta_{D}\equiv\delta_{D}(m_{+}^{2},m_{-}^{2}) ~-~ \delta_{D}(m_{-}^{2},m_{+}^{2})$, and $\delta_{D}(m_{+}^{2},m_{-}^{2})$ is the strong phase of $f_{D}(m_{+}^{2},m_{-}^{2})$. Hence, knowledge of $\Delta\delta_{D}$ is essential for the determination of $\gamma$ in $B^-\to DK^-$ decays. 

In the literature, both model-dependent and model-independent BPGGSZ methods are used. In the model-dependent approach, the $D^0$ amplitude
is obtained using a flavor-tagged $D^{0}$ meson sample selected from the $D^{*\pm} \to D^0(K^{0}_{\rm S}K^{+}K^{-})\pi^{\pm}$ decay, which is fit to an amplitude model describing the decay of $D^{0} \to K^{0}_{\rm S}K^{+}K^{-}$~\cite{babarmodel} to determine $f_D(m_+^2,m_-^2)$. The amplitude model is then used in an unbinned likelihood fit to the $B$-meson data sample to determine $\gamma$, $\delta_B$, 
and $r_B$. However, this method results in a model-dependent systematic uncertainty on the measured value of $\gamma$ which is difficult to quantify~\cite{battaglieri}. These model-dependent uncertainties have been estimated to lie between $3^\circ$ to $9^\circ$~\cite{amo,poluektov}, which limits the precision on $\gamma$ that future measurements performed with much larger $B$-meson data samples \cite{LHCb-upgrade,BelleII} can obtain. 

An alternative method of measuring $\gamma$ is in a model-independent manner that relies on defining a number of bins in the $D^0\to K^{0}_{\rm S}K^{+}K^{-}$ Dalitz plot \cite{giri}. This approach determines $\gamma$ from the measured rate in each bin of the Dalitz plot, rather than fitting the Dalitz plot distribution to an amplitude model. The method requires information about $\Delta\delta_{D}(m_{+}^{2},m_{-}^{2})$ in each bin, which is accessible at the $\psi(3770)$ resonance
by exploiting the quantum coherence of the $D^{0}\bar{D}^{0}$ pair produced in $\psi(3770)$ decays. The advantage of this method is that the hard-to-quantify systematic uncertainty related to the model assumption is replaced by the uncertainty on the binned strong-phase parameters of the $D$ decay mode. These strong-phase parameter uncertainties are statistically dominated, and thus well understood. The major disadvantage of the model-independent method is the inevitable loss of information that arises from binning, which reduces the statistical sensitivity of the $\gamma$ measurement by approximately 80\% compared to the model-dependent method \cite{jim}.

The remainder of this paper is structured as follows. In Sec.~\ref{sec:formalism} we define the formalism used to measure the strong-phase parameters with $\psi(3770)$ data. We explain the Dalitz-plot binning in Sec.~\ref{sec:dalitzbinning}. In Sec.~\ref{sec:Bes3detectorandsimulation} we outline the features of the BESIII detector and the simulation techniques used in the analysis. We describe the event-selection criteria and the procedure for estimating the data yields in Secs.~\ref{sec:Eventselection} and \ref{sec:yieldestimation}, respectively. In Sec.~\ref{sec:dalitzplot} we explain the procedure for estimating the bin yields, including the various corrections applied. We describe the extraction of strong-phase parameters and the calculation of systematic uncertainties in Secs.~\ref{sec:cisiextraction} and \ref{sec:systematics}, respectively. We present a discussion on the impact of these results on $\gamma$ in Sec.~\ref{sec:impact}. In Sec.~\ref{sec:summary} we give the conclusion and outlook.

\section{\boldmath Formalism}\label{sec:formalism}
The model-independent method \cite{giri} for a three-body $D$ decay is implemented as follows. The entire Dalitz plot is divided into 2$\mathcal{N}$ bins, with $\mathcal{N}$ bins symmetrically placed on either side of the $m^{2}_{+} = m^{2}_{-}$ line.  We follow the convention in which bins with $m_+^2\geq m_-^2$ are labelled with $i$ and bins with $m_+^2<m_-^2$ are labelled with $-i$. Thus, the $2\mathcal{N}$ bins are assigned labels from $\mathcal{-N}$ to $\mathcal{N}$ excluding zero.
The interchange of the Dalitz plot variables $m^{2}_{+}\leftrightarrow m^{2}_{-}$ corresponds to the interchange of positions of the bins $i\leftrightarrow -i$. In order to extract the strong-phase difference parameters, we need to determine the yield in each bin for flavor-, $CP$- and mixed-$CP$ tagged $D\to K^0_{\rm S} K^+ K^-$ decays. The number of flavor-tagged $D^{0}\to K_{\rm S}^{0} K^{+}K^{-}$ decays $K_i$ in the $i$th bin of the Dalitz plot is defined as
\begin{equation}\label{eqn:relationkiandti}
K_{i} = a_{D}\int_i \left|f_{D}\left(m_{+}^{2},m_{-}^{2}\right)\right|^{2}~ {\rm d}m^{2}_{+} ~{\rm d}m^{2}_{-} = a_{D}F_{i},
\end{equation}
where $a_{D}$ is a normalization factor equal to the total number of $D^{0}\rightarrow K^{0}_{\mathrm{S}}K^{+}K^{-}$ decays in the flavor-tagged charm sample, $F_{i}$ is the fraction of $D^{0}\to K_{\rm S}^{0} K^{+}K^{-}$ decays in the $i$th bin, and the integral is over the $(m_{+}^2,m_{-}^2)$ region defined by the $i$th bin. Here and elsewhere the values of $K_{(-)i}$ are corrected for efficiency and also for the presence of any doubly Cabibbo-suppressed (DCS) component (see Sec.~\ref{sec:binyieldandcorrection}). We assume $f_D(m_+^2,m_-^2)$ has been normalized such that 
\begin{equation}
 \int \left|f_D\left(m_{+}^2,m_{-}^2\right)\right|^2~{\rm d}m_{+}^2\rm{d}m_{-}^2 = 1\;,
\end{equation}
where the integral is over the whole Dalitz plot.
For each bin the interference between $D^{0}$ and $\bar{D}^{0}$ decays can be parameterized by two variables $c_{i}$ and $s_{i}$, which are the amplitude-weighted averages of $\cos{ \Delta\delta_{D}}$ and $\sin{\Delta\delta_{D}}$, defined as: 
\begin{eqnarray}
c_{i} \equiv \frac{1}{\sqrt{F_{i}F_{-i}}}&\bigintsss\limits_{i}&|f_{D}(m_{+}^{2},m_{-}^{2})||f_{D}(m_{-}^{2},m_{+}^{2})| \nonumber \\
&\times&  \mathrm{cos}[\Delta\delta_{D}(m_{+}^{2},m_{-}^{2})] dm_{+}^{2}dm_{-}^{2},
\label{eq:cieqn}
\end{eqnarray}
and 
\begin{eqnarray}
s_{i} \equiv \frac{1}{\sqrt{F_{i}F_{-i}}}&\bigintsss\limits_{i}&|f_{D}(m_{+}^{2},m_{-}^{2})||f_{D}(m_{-}^{2},m_{+}^{2})| \nonumber \\ 
&\times& \mathrm{sin}[\Delta\delta_{D}(m_{+}^{2},m_{-}^{2})] dm_{+}^{2}dm_{-}^{2}.
\label{eq:sieqn}
\end{eqnarray}
From Eqs.~(\ref{eq:cieqn}) and ~(\ref{eq:sieqn}) it is evident that $c_{i} = c_{-i}, s_{i} = -s_{-i}$ and $c_{i}^{2}+s_{i}^{2} \leq 1$. The equality is satisfied only if $f_{D}$ is constant throughout the bin. Thus the yield of $B^{\pm}$ decays in the $i$th bin, $N_{i}$, is obtained by integrating Eq.~(\ref{eq:bdecay_rate}), which results in 
\begin{equation}\label{eq:bdecayrate}
N_{i}^{\mp} \propto K_{\pm i} + r_{B}^{2}K_{\mp i} + 2\sqrt{K_{i}K_{-i}}(x_{B\mp}\cdot c_{i} + y_{B\mp}\cdot s_{i}),
\end{equation}
where $x_{B\mp} \equiv r_{B}\mathrm{cos}(\delta_{B}\mp\gamma)$, $y_{B\mp} \equiv r_{B}\mathrm{sin}(\delta_{B}\mp\gamma)$ and $r_{B}^{2} = x_{B\mp}^{2} + y_{B\mp}^{2}$. 
A maximum likelihood fit to binned $B^-\to DK^-$ decay yields, using Eq.~(\ref{eq:bdecayrate}) as a probability density function with externally measured values of $c_{i}$ and $s_{i}$ as inputs, then allows $\gamma$ to be determined along with $r_{B}$ and $\delta_{B}$. 

We now describe how $\psi(3770)$ data are used to determine the values of $c_{i}$ and $s_{i}$. The $D^{0}\bar{D}^{0}$ pair from the decay of the $\psi(3770)$ (or if directly produced from the virtual photon in an $e^+e^-$ annihilation) is in a $C$-odd eigenstate, as long as there are no additional particles in the final state. This quantum correlation between the mesons leads to the total $D^0\bar{D}^0$ decay rate being sensitive to the strong-phase difference between the $D^0$ and $\bar{D}^0$ amplitudes. 
For example, the decay of one $D$ to a $CP$-even eigenstate fixes the other $D$
to the $CP$-odd admixture of $\left(D^0-\bar{D}^0\right)/\sqrt{2}$. Hence, if the other
$D$ decays to $K_{\rm S,L}K^+K^-$, the total rate will be sensitive to the
interference between the $D^0$ and $\bar{D}^0$ amplitudes and the strong-
phase parameters.
Generally, this interference affects the decays of one $D$ in combination with the other.
If only one $D$ meson is reconstructed, leaving the companion $D$ meson to decay to any final state, the decay rate is largely insensitive to the effects of quantum correlations; we refer to the reconstructed samples of such events as single-tag (ST) decays.
If both $D$ mesons are required to be in specific final states, the rates can be significantly enhanced or suppressed in the quantum-correlated events compared to the expected rate if the decays are uncorrelated; we refer to the reconstructed samples of such events as double-tag (DT) decays. Hereafter, all the $D$ decay final states, except the signal mode $K^0_{\mathrm{S}, \mathrm{L}} K^+K^-$, are referred to as ``tags''. 

The $D\to K^{0}_{\rm S}K^{+}K^{-}$ decay amplitude from a $CP$ eigenstate is
\begin{equation}
    f_{\pm}\left(m^{2}_{+}, m^{2}_{-}\right) = \frac{1}{\sqrt{2}}\left[f_{D}\left(m^{2}_{+}, m^{2}_{-}\right)\pm f_{D}\left(m^{2}_{-}, m^{2}_{+}\right)\right]\;,
\end{equation}
where $+~(-)$ indicates a $CP$-even ($CP$-odd) state. 
Therefore, the expected number of events $\langle M^{\pm}_{i} \rangle$ in the $i$th bin of a sample that has been tagged with a decay that has a $CP$-even fraction $F_+$ is  
\begin{equation}
\langle M^{\pm}_{i} \rangle = \epsilon_{{\rm DT},i} \frac{S_{\pm}}{2S_{f}}\left(K_{i} - 2c_{i}(2F_{+} - 1) \sqrt{K_{i}K_{-i}} + K_{-i}\right),
\label{eq:Mieqn}
\end{equation}
where $S_{\pm}$ $(S_{f})$ are the efficiency-corrected single-tag yields of the $CP$-eigenstate (flavor) modes
used in the analysis and $\epsilon_{{\rm DT},i}$ is the DT efficiency in the $i$th bin. The value of $F_{+}$ is equal to 1 (0) for a pure $CP$-even ($CP$-odd) tag mode. We refer to modes with intermediate values of $F_{+}$ as quasi-$CP$ tags. The values of $c_{i}$ alone can be extracted using Eq.~(\ref{eq:Mieqn}). The relation $s_{i} = \sqrt{1-c_{i}^{2}}$ is a good approximation only for $\mathcal{N}>200$~\cite{bondar}, which is not feasible with the available data sample. However, analysing $D \to K_{\rm S}^{0}K^{+}K^{-}$ decays tagged by $D \to K_{\rm S}^{0}h^{+}h^{-}$ ($h$ = $\pi$, $K$)
decays gives access to both $c_{i}$ and $s_{i}$.
The amplitude of the $D^{0}\bar{D}^{0}$ pair produced by the $\psi(3770)$ decaying to $K_{\rm S}^{0}K^{+}K^{-}$ and $K_{\rm S}^{0}h^{+}h^{-}$ is 
\begin{widetext}
\begin{equation}
f_{D}(m^{2}_{+},m^{2}_{-},\overline{m}^{2}_{+},\overline{m}^2_{-}) = \frac{f_{D}(m^{2}_{+},m^{2}_{-}) f_{D}(\overline{m}^{2}_{-},\overline{m}^{2}_{+}) - f_{D}(\overline{m}^{2}_{+},\overline{m}^{2}_{-})f_{D}(m^{2}_{-},m^{2}_{+})}{\sqrt{2}}\;,
\end{equation}
\end{widetext} 
where $(\overline{m}_{+}^2,\overline{m}_{-}^2)$ are the Dalitz plot coordinates corresponding to the phase space of the $K_{\rm S}^{0}h^{+}h^{-}$ decay. The expected event rate in which one $D$ decays in the region of phase space defined by the $i$th bin of the $D\to K_{\rm S}^{0}K^{+}K^{-}$ Dalitz plot and the other $D$ in the region of phase space defined by the $j$th bin of the $D\to K_{\rm S}^{0}h^{+}h^{-}$ Dalitz plot can be written as
\begin{eqnarray}
\langle M_{ij} \rangle &=& \epsilon_{{\rm DT},ij} \frac{N_{D^{0}\bar{D}^{0}}}{2 S_{f}^{2}}\Big( K_{i}K_{-j} + K_{-i}K_{j}\nonumber \\ 
&& - 2\sqrt{K_{i}K_{-j}K_{-i}K_{j}} ~(c_{i}c_{j}+s_{i}s_{j})\Big),
\label{eq:mijeqn}
\end{eqnarray}
where $N_{D^{0}\bar{D}^{0}}$ is the number of $D^{0}\bar{D}^{0}$ pairs in the $\psi(3770)$ data sample and $\epsilon_{{\rm DT},ij}$ is the DT efficiency in the $i$th and $j$th pair of bins. The two-fold ambiguity in the sign of $s_{i}$ can be resolved using weak amplitude-model assumptions. Note that Eq.~(\ref{eq:Mieqn}) is symmetric under the interchange of $i\leftrightarrow -i$ and Eq.~(\ref{eq:mijeqn}) is symmetric under the interchange of pair, $(i,j)\leftrightarrow(-i,-j)$ and $(i,-j)\leftrightarrow(-i,j)$. No such symmetry exists for the values of $K_{i}$ because $f_{D}(m^{2}_{+},m^{2}_{-})\neq f_{D}(m^{2}_{-},m^{2}_{+})$. 

Ignoring the very low level of $CP$ violation in the neutral kaon system, the $K^{0}$ state is an equal admixture of $K^{0}_{\rm S}$ and $K^{0}_{\rm L}$ states. Therefore, in the decays of correlated $D^{0}\bar{D}^{0}$ pairs we expect a significant fraction of the ${D}$ mesons to decay to the $K^{0}_{\rm L}K^{+}K^{-}$ final state as well. Although so far $\gamma$ has only been determined using $D\to K^{0}_{\rm S}h^{+}h^{-}$ decays, the decay $D\to K^{0}_{\rm L}K^{+}K^{-}$ has a close connection with $D\to K^{0}_{\rm S}K^{+}K^{-}$ that can be exploited to improve the precision with which $c_{i}$ and $s_{i}$ are determined. In the absence of $CP$ violation, $CP|K^{0}_{\rm S}\rangle = |K^{0}_{\rm S}\rangle$ and $CP|K^{0}_{\rm L}\rangle = -|K^{0}_{\rm L}\rangle$. Hence $K^{0}_{\rm L}K^{+}K^{-}$ has opposite $CP$ to $K^{0}_{\rm S}K^{+}K^{-}$. We define 
the decay amplitude for $D^{0}\to K^{0}_{\rm L}K^{+} K^{-}$ [$\bar{D}^{0}\to K^{0}_{\rm L}K^{+} K^{-}$] as $f^{\prime}_{D}\left(m_+^{2},m_{-}^{2}\right)$ [$f^{\prime}_{\bar{D}}\left(m_+^{2},m_{-}^{2}\right)$] such that 
\begin{equation}
f^{\prime}_{\bar{D}}\left(m^{2}_{+},m^{2}_{-}\right) = -f^{\prime}_{D}\left(m^{2}_{-},m^{2}_{+}\right)\;.
\end{equation}
Therefore, the number of events in the $i$th bin of a $CP$- or quasi-$CP$ tagged $D\to K^{0}_{\mathrm{L}}K^{+}K^{-}$ sample is 
\begin{equation}
\langle M_{i}^{'\pm} \rangle = \epsilon_{{\rm DT},i}^{\prime} \frac{S_{\mp}}{2S'_{f}}\left(K'_{i} + 2c'_{i}(2F_{+} - 1)\sqrt{K'_{i}K'_{-i}} + K'_{-i}\right)\;,
\label{eq:Miprimeeqn}
\end{equation}
where $K_{i}^{\prime}$ is defined in analogous fashion to $K_{i}$ (see Eq.~(\ref{eqn:relationkiandti})). 
Furthermore, the expected event rate in the $i$th bin of the $D\to K_{\rm L}^{0}K^{+}K^{-}$ Dalitz plot and the $j$th bin of the $D\to~K_{\rm S}^{0}h^{+}h^{-}$ Dalitz plot can be written as
\begin{eqnarray}
\langle M'_{ij} \rangle &=& \epsilon_{{\rm DT},ij}^{\prime}\frac{N_{D\bar{D}}}{2 S_{f}S'_{f}}\Big(K_{i}K'_{-j} + K_{-i}K'_{j} \nonumber \\
&+&  2\sqrt{K_{i}K'_{-j}K_{-i}K'_{j}} ~(c'_{i}c_{j}+s'_{i}s_{j})\Big).
\label{eq:Mijprimeeqn}
\end{eqnarray}
 The  symmetries between the exchange of coordinates in the cases of $M_{i}$ and $M_{ij}$ are also present for $M^{\prime}_{i}$ and $M_{ij}^{\prime}$. In general ($c_{i},s_{i})\neq (c_{i}^{\prime},s_{i}^{\prime})$ because $f_{D}^{\prime}\left(m^2_{+},m^2_{-}\right)\neq f_{D}\left(m^2_{+},m^2_{-}\right)$. In order to improve the precision of the extracted values of $c_{i}$ and $s_{i}$ constraints are imposed on the difference $\Delta c_{i}=c_{i}^{\prime}-c_{i}$ and  $\Delta s_{i}=s_{i}^{\prime}-s_{i}$; these constraints are explained in Sec.~\ref{sec:cisiextraction}.

\section{\boldmath Binning of the  $D^{0}\to K^{0}_{\rm S}K^{+}K^{-}$  Dalitz plot}\label{sec:dalitzbinning}
\begin{figure*}[ht!] \centering
\includegraphics[width=7.0in,height=7.0in,keepaspectratio]{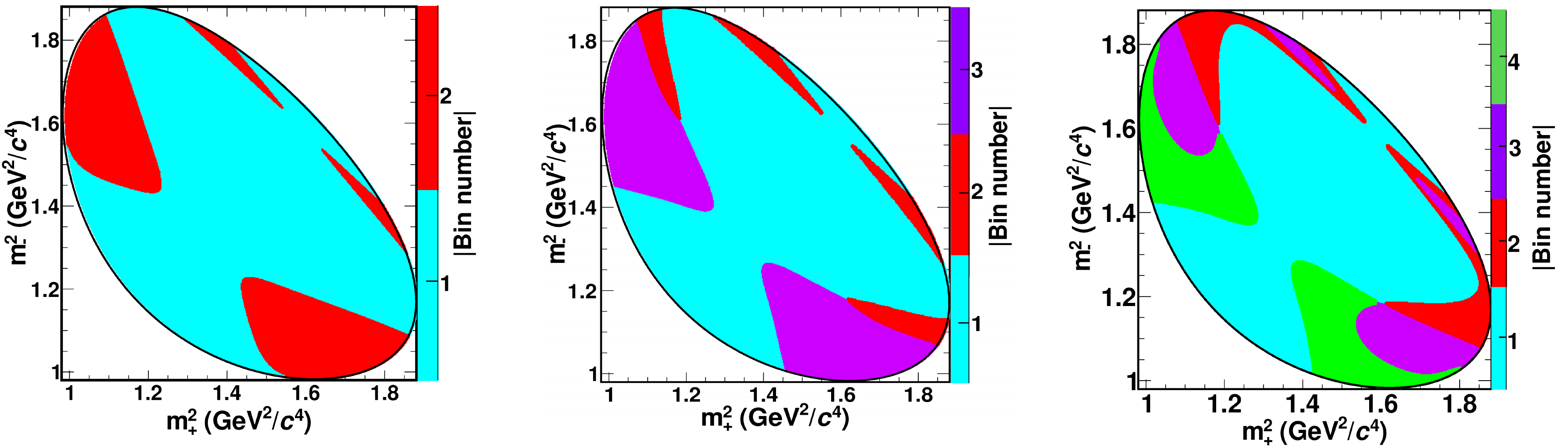}
\caption{Equal-$\Delta\delta_{D}$ binning of $D^{0}\to K^{0}_{\rm S}K^{+}K^{-}$ phase-space based on the BaBar model~\cite{babarmodel} for $\mathcal{N}=2$ (left), $\mathcal{N}=3$ (middle) and $\mathcal{N}=4$ (right) bins. The color scale represents the absolute value of the bin number and the black curve represents the kinematic boundary of the Dalitz plot.}
\label{fig:kskkbins_model}
\end{figure*}

All the relations given in Sec.~\ref{sec:formalism} are independent of the shape of the Dalitz plot bins. The original proposal \cite{giri} was to divide the Dalitz plot into rectilinear bins. The reduction in sensitivity of such an approach compared to an unbinned analysis is about 30\% even with 20 bins \cite{bondar}. The sensitivity of the model-independent method as a function of the bin shape is discussed in Ref.~\cite{bondar}; this paper concludes that binning schemes that minimise the variations of $\Delta\delta_{D}$ within each Dalitz plot bin give significantly improved statistical sensitivity compared to the rectilinear binning. An amplitude model can be used to guide the definition of bin boundaries in order to minimize the  $\Delta\delta_D$ variation. The number of bins that can be used in the analysis is restricted by the available statistics in either the $\psi(3770)$ or $B$-decay data samples. Since the use of an amplitude model is only to define the bin shapes, the model neither leads to any bias nor introduces any model-dependent uncertainties on the measurement of $\gamma$. However, a model that poorly describes the phase variation of the amplitude over the Dalitz plot may lead to a lower than expected statistical sensitivity to $\gamma$. 

In the current analysis we employ an amplitude model for $D^0 \to K^{0}_{\rm S}K^{+}K^{-}$ decays developed by the BaBar Collaboration~\cite{babarmodel} to define the bin shapes. Our choice of model and bin definitions is consistent with the previous measurement~\cite{jim}. The amplitude model is constructed in the isobar formalism, where the amplitude at a phase-space point is defined as a coherent sum of two-body amplitudes and a non-resonant amplitude. There are eight intermediate resonances used in the model. The $a_{0}(980)^{0}$ and $a_{0}(980)^{\mp}$ resonances are modelled by the Flatt\'e parameterization~\cite{flatte}, while all other resonances are parameterized by Breit-Wigner line shapes.
The model-based lookup table (LUT) containing the moduli and phases of the $D^{0}\to K_{\rm S}^{0}K^{+}K^{-}$ amplitudes at different phase points ($m^{2}_{-}, M^{2}_{K^{+}K^{-}}$) was supplied by the authors of Ref.~\cite{babarmodel}. The granularity of the ($m^{2}_{-}, M^{2}_{K^{+}K^{-}}$) grid in the LUT is 0.00179 GeV$^{2}/c^{4}\times$ 0.00536 GeV$^{2}/c^{4}$. Based on the LUT, the values of $\Delta\delta_{D}$ at a position ($m^{2}_{+},m^{2}_{-}$) in the phase-space are calculated. Half of the Dalitz plot, $m_{+}^{2} < m_{-}^{2}$, is divided into equally-spaced regions (bins) of $\Delta\delta_{D}$ satisfying the condition 
\begin{equation}\label{eq:equaldeltaD}
2\pi(i - 3/2)/{\mathcal{N}} \leq \Delta\delta_{D}(m_{+},m_{-}) < 2\pi(i - 1/2)/{\mathcal{N}},
\end{equation}
as shown in Fig.~\ref{fig:kskkbins_model} for $\mathcal{N}=2,~3$ and 4. Here $i=1,~2,...,\mathcal{N}$ are the bin numbers. The bins in the region $m_{+}^{2} > m_{-}^{2}$ are defined symmetrically. The class of binning defined by Eq.~(\ref{eq:equaldeltaD}) is referred to as the ``equal-$\Delta\delta_{D}$" binning scheme. A smaller number of bins is the best choice to measure $c_{i}$ and $s_{i}$ precisely, but this will potentially reduce the sensitivity to $\gamma$. On the other hand, a larger number of bins provides increased sensitivity to $\gamma$, because it is a better approximation to the unbinned method. Keeping this trade-off in mind, we perform the analysis with $\mathcal{N} = 2,~3$ and 4 bins. Binning the Dalitz plot
with $\mathcal{N} > 4$ is not yet feasible with the size of the sample of $\psi(3770)$ data collected by BESIII; the fit to determine $c_i$ and $s_i$ described in Sec. ~\ref{sec:cisiextraction} fails with $\mathcal{N}>4$. 

In order to ascertain the quality of the binning, a figure-of-merit based on the ratio of statistical sensitivity of the binned to the unbinned approach, known as the binning quality factor, $\mathcal{Q}$, is defined in Ref.~\cite{bondar}. The predicted values of $\mathcal{Q}$ for this model are determined to be 0.771, 0.803 and 0.822 for $\mathcal{N} = 2,~3$ and 4 bins, respectively~\cite{jim}. The measured values were 0.94$^{+0.16}_{-0.06}$, 0.87$^{+0.14}_{-0.06}$ and 0.94$^{+0.21}_{-0.06}$ for $\mathcal{N} = 2,~3$ and 4 bins respectively~\cite{chris_thesis}. Since these values are close to one it implies that the loss of sensitivity due to the current bin definitions is small. An optimal binning scheme, which accounts for the distribution of the $B$-meson data sample across the Dalitz plot, as well as the $\Delta\delta_D$ variation, is found to give negligible improvement to the projected sensitivity compared to the ``equal-$\Delta\delta_D$" binning \cite{jim}; hence, it is not pursued further.

\section{\boldmath BESIII detector and event generation}\label{sec:Bes3detectorandsimulation}

We analyse an $e^{+}e^{-}$ collision data sample produced by the Beijing Electron Positron Collider II (BEPCII), which corresponds to an integrated luminosity of 2.93~fb$^{-1}$~\cite{bepcluminosity}, collected by the BESIII detector at a center-of-mass energy of $\sqrt{s} = 3.773$ GeV.
The BESIII experiment is a general purpose solenoidal detector with a geometrical acceptance of 93\% of the 4$\pi$ solid angle. It has a He-gas-based  multilayer drift chamber (MDC) for measuring the momentum and specific ionization loss ($dE/dx$) of the charged particles, a plastic-scintillator-based time-of-flight (TOF) measurement system for the identification of charged particles, and an electromagnetic calorimeter (EMC) consisting of CsI(Tl) crystal, which is used to measure the energy of the neutral showers and identify electrons. The detector is encapsulated in a magnetic field of 1~T provided by a superconducting solenoid. A resistive-plate-chamber-based muon counter is interleaved between the flux-return yoke of the magnet. The MDC has a transverse-momentum resolution of 0.5\% at 1 GeV/$c$. The time resolution of the TOF is about 80~ps in the barrel region and 110~ps in the endcap region, enabling a 2$\sigma$ $K/\pi$ separation up to a momentum of 1~GeV/$c$. The energy resolution of the EMC for 1~GeV photons is about 2.5\% in the barrel region and 5\% in the endcap regions. More details about the BESIII detector can be found in Ref.~\cite{bes3detector}.

Simulated samples produced with the {\sc
geant4}-based~\cite{geant4} Monte Carlo (MC) package, which
includes the geometric description of the BESIII detector and the
detector response, are used to determine the detection efficiency
and to estimate the backgrounds. The simulation includes the beam-energy spread and initial-state radiation (ISR) in the $e^+e^-$
annihilations modelled with the generator {\sc
kkmc}~\cite{kkmc}. The inclusive MC samples consist of the production of $D\bar{D}$
pairs, the non-$D\bar{D}$ decays of the $\psi(3770)$, the ISR
production of the $J/\psi$ and $\psi(3686)$ states, and the
continuum processes incorporated in {\sc kkmc}~\cite{kkmc}.
The known decay modes are modelled with {\sc
evtgen}~\cite{evtgen} using branching fractions taken from the
Particle Data Group~\cite{pdg}, and the remaining unknown decays
from the charmonium states with {\sc
lundcharm}~\cite{lundcharm}. The final state radiation (FSR)
from charged final-state particles is incorporated with the {\sc
photos} package~\cite{photos}. The simulation of  quantum-correlations in the process $\psi(3770)\to D^0\bar{D}^0$ is done outside the {\tt EVTGEN} framework, using an algorithm developed by the CLEO Collaboration~\cite{danthesis}. The effective integrated luminosity of the generated $D^{0}\bar{D}^{0}$ sample is about ten times that of the data. For the efficiency determination we use signal MC samples. Signal MC samples consist of $D^{0}\to S_{\rm tag},\bar{D}^{0}\to X$ decays for the reconstruction of STs and $D^{0}\to K^{0}_{\rm S,L}K^{+}K^{-}, \bar{D}^{0}\to S_{\rm tag}$ decays for the reconstruction of DTs, where $S_{\rm tag}$ is a tag final state and $X$ is any inclusive final state. Each signal MC sample corresponds to a specific ST or DT decay mode studied in this paper and contains $2\times 10^{5}$ events.

\section{\boldmath Event selection}\label{sec:Eventselection}
In this section we initially describe the requirements for selecting the reconstructed particles that are combined to form the final states of interest. Then we present the selection criteria of fully reconstructed tag modes and partially reconstructed tag modes in Secs.~\ref{sec:fullselection} and~\ref{sec:partialselection}, respectively.

Table \ref{tab:tagmodes} summarizes the set of tag modes used to reconstruct $D^{0}$ final states. The decay channels are split into five categories: Signal, Flavored, Mixed $CP$, $CP$-odd and $CP$-even. A highlight of this analysis is that the quasi-$CP$ mode $D\to \pi^{+}\pi^{-}\pi^{0}$, which has a large branching fraction, is used for the first time for the strong-phase measurements in the $D\to K^{0}_{\rm S, L}K^{+}K^{-}$ analysis. The $F_{+}$ value of $\pi^{+}\pi^{-}\pi^{0}$ is measured in Ref.~\cite{minakshi} and the mode is found to be overwhelmingly $CP$-even. Hence in this analysis we treat $\pi^{+}\pi^{-}\pi^{0}$ as a $CP$-even tag taking into account its $F_{+}$ value.
In the analysis, daughter particles are reconstructed as: $K^{0}_{\rm S}\rightarrow\pi^{+}\pi^{-}$, $\eta\rightarrow\gamma\gamma$, $\pi^{0}\rightarrow\gamma\gamma$, $\omega\rightarrow\pi^{+}\pi^{-}\pi^{0}$, $\eta'\rightarrow\pi^{+}\pi^{-}\eta$. In this section we will describe the selection criteria implemented to reconstruct these final states. 
\begin{table}[ht!] 
	\centering
	\caption{$D^{0}$ decays used in this analysis.}	
  \begin{tabular}{l c }  
  \hline
  \hline
	Type & Tag modes\\
	\hline
	Signal & $K_{\rm S}^{0}K^{+}K^{-}$, $K_{\rm L}^{0}K^{+}K^{-}$\\
	Flavored & $K^{-}\pi^{+}$, $K^{-}\pi^{+}\pi^{0}$, $K^{-}e^{+}\nu_{e}$ \\	
	Mixed $CP$ & $K^{0}_{\rm S}\pi^{+}\pi^{-}$, $K^{0}_{\rm L}\pi^{+}\pi^{-}$ \\
	$CP$-odd & $K_{\rm S}^{0}\pi^{0}$, $K_{\rm S}^{0}\eta$, $K^{0}_{\rm S}\eta'$, $K_{\rm S}^{0}\omega$\\
	$CP$-even	& $K^{+}K^{-}$, $\pi^{+}\pi^{-}$, $\pi^{+}\pi^{-}\pi^{0}$, $K_{\rm S}^{0}\pi^{0}\pi^{0}$,\\
	   & $K_{\rm L}^{0}\pi^{0}$, $K_{\rm L}^{0}\eta$, $K^{0}_{\rm L}\eta'$, $K_{\rm L}^{0}\omega$ \\
	   \hline
	   \hline
\end{tabular}
\label{tab:tagmodes}
\end{table} 

For the charged tracks the polar angle $\theta$ is required to be within the MDC acceptance, which is $\left|\rm cos~\theta \right|$ $<$ 0.93. The distance of closest approach of a primary track from the interaction region is required to be less than 10~cm in beam direction and less than 1~cm in the plane perpendicular to the beam direction to remove tracks not originating from $e^{+}e^{-}$ collisions. For neutral showers the energy deposited in the EMC is required to be larger than 0.025~GeV in the barrel region $\left(\left| \rm cos ~\theta \right|< 0.8\right)$ and larger than 0.050~GeV in the endcap region $\left(0.86< \left| \rm cos ~\theta \right| < 0.92\right)$, which reduces the effect of electronic noise and deposits resulting from beam-related backgrounds. Moreover, the angle between the position of the shower and any extrapolated charged track in the EMC must be greater than $10^\circ$ to reduce the number of showers related to charged tracks. Furthermore, we require the time of the shower to be less than 700~ns after the event start-time to further suppress fake photons associated with electronic noise and beam backgrounds.

The particle identification (PID) is performed by combining the $dE/dx$ information from the MDC as well as the time-of-flight of the charged particle. The likelihoods for the kaon hypothesis $\mathcal{L}_{K}$ and pion hypothesis $\mathcal{L}_{\pi}$ are calculated. Tracks satisfying the condition $\mathcal{L}_{K} > \mathcal{L}_{\pi}$ are identified as kaons and vice versa for pions. For electrons the PID is performed by defining a likelihood based on information about $dE/dx$ in the MDC, time-of-flight and deposited energy and shape of the electromagnetic shower from the EMC.
The track is identified as an electron if $\mathcal{L}_{e}/(\mathcal{L}_{e}+\mathcal{L}_{K}+\mathcal{L}_{\pi}) > 0.8$ and $\mathcal{L}_{e} > $ 0.001, where $\mathcal{L}_{e}$ is the likelihood of the electron hypothesis. 

A $K^{0}_{\rm S}$ candidate is formed by considering a pair of intersecting oppositely charged tracks. These tracks are not subject to any track quality requirement or PID. The closest approach of these tracks to the interaction point is required to be less than 20 cm along the beam direction with no requirement in the transverse direction. A secondary vertex fit is performed to form the $K^{0}_{\rm S}$ vertex, and candidates with $\chi^{2} < $ 100 are selected. The updated four momenta after the secondary vertex fit are used later in this analysis. The mass of a $K^{0}_{\rm S}$ candidate is required to be within the range (0.487, 0.511)~GeV/$c^{2}$. In order to suppress combinatorial backgrounds from two pions that are not from a true $K^{0}_{\rm S}$, the flight significance, $L/\sigma_{L}$, is required to be greater than two, where $L$ is the flight length and $\sigma_{L}$ is the uncertainty in $L$ from the secondary vertex fit.

Both $\pi^{0}$ and $\eta$ candidates are reconstructed from a pair of photons, where at least one of the photons must be reconstructed in the barrel region; this requirement reduces combinatorial backgrounds that arise from the large number of showers in the endcap region that are related to beam backgrounds. The invariant mass of the two photon candidates must be in the range (0.110,~0.155)~GeV$/c^{2}$ or (0.480,~0.580)~GeV$/c^{2}$ for $\pi^0$ and $\eta$ candidates, respectively. In order to improve the momentum resolution, a kinematic fit of the two photons is performed with their invariant mass constrained to the nominal mass of $\pi^{0}$ or $\eta$~meson taken from the PDG~\cite{pdg}. Only $\pi^{0}$ and $\eta$ candidates with $\chi^{2}$ $<$ 20 are selected. The improved values of the momenta are used later in the analysis. For $\omega$ candidates the invariant mass of the $\pi^{+}\pi^{-}\pi^{0}$ combination is required to be within the range (0.760, ~0.805)~GeV/$c^{2}$ and for $\eta^{\prime}$ candidates the invariant mass of the $\pi^{+}\pi^{-}\eta$ combination is required to be within the range (0.938,~ 0.978)~GeV/$c^{2}$. All the invariant mass intervals described correspond to approximately $\pm3$ times the standard deviation about the mean of the reconstructed distribution.  

\subsection{\boldmath Selection of fully reconstructed tags} \label{sec:fullselection} 
Fully reconstructed tags are decay modes that do not contain an undetected particle in the final state. Before describing the kinematic variables used to select fully reconstructed tags, we introduce two additional vetoes that remove specific backgrounds to certain tag modes. The first veto is to suppress backgrounds arising from cosmic rays and lepton-pair events in the ST reconstruction of the two-body decay channels $K^{+}K^{-}$, $\pi^{+}\pi^{-}$ and $K^{\pm}\pi^{\mp}$. 
Here, we reject events in which the two charged tracks that reconstruct the ST candidate are consistent with being an $e^{+}e^{-}$ or $\mu^+\mu^-$ pair.
%
In addition, to suppress cosmic muons, we reject events where the time-of-flight difference between the two tracks is greater than 5~ns. Further, an event that has neither an EMC shower with an energy greater than 50 MeV nor an additional charged track in the MDC is rejected.
The second veto is to remove the $CP$-odd $K^0_{\rm S}\pi^0$, $K^0_{\rm S}\to \pi^{+}\pi^{-}$ background to the predominantly $CP$-even $\pi^{+}\pi^{-}\pi^{0}$ tag mode; here we reject events that satisfy the condition ~$|M_{\pi^{+}\pi^{-}} - m_{K^{0}_{\rm S}}| < 0.018$ GeV$/c^{2}$, where $m_{K^{0}_{\rm S}}$ refers to the nominal mass of the $K^{0}_{\rm S}$ meson given in Ref.~\cite{pdg}.

 For all fully reconstructed tag modes, the selected final-state particles are combined to reconstruct the $D$ decay. Since the $D\bar{D}$ pair production occurs at the $\psi(3770)$ resonance, there are no additional particles in the final state, so the energy of each $D$ meson is equal to $\sqrt{s}/2$. Thus, with a well measured beam energy $E_{\rm beam}$ ($= \sqrt{s}/2$) we define two quantities to reconstruct the $D$ candidates: the energy difference,
\begin{equation} 
\Delta E \equiv \sum_{i} E_{i} - E_{\rm beam} \mathrm{,} 
\end{equation}
and the beam-constrained mass,
\begin{equation}   
M_{\rm BC} \equiv \frac{1}{c^2}\sqrt{E^{2}_{\rm beam} - \left|\sum_{i} \textbf{p}_{i}c\right|^{2}} .
\end{equation}
Here $ E_{i} $ and  $\textbf{p}_{i}$  are the energies and momenta of the \textit{D} decay products in the center-of-mass frame. Properly reconstructed candidates will peak at zero in the $\Delta E$ distribution and at the nominal mass of the $D^{0}$ meson~\cite{pdg} in the $M_{\rm BC}$ distribution. For all the reconstructed final states mode-dependent criteria are applied to the $\Delta E$ distribution to reduce the level of combinatorial backgrounds.
 The $\Delta E$ distribution is fit with a combination of a double-Gaussian function and a polynomial to describe the signal and background, respectively. The value of $\Delta E$ is required to be within the range $\pm 3\sigma$  [$(-4\sigma,3\sigma)$] from the mean of the signal distribution for modes without [with] a $\pi^{0}$ in the final state. Here $\sigma$ is the total width of the $\Delta E$ signal shape. If multiple ST candidates are reconstructed in an event, the candidate with the minimum value of $\left|\Delta E\right|$ is selected. If multiple DT candidates are selected, the candidate with a value $\overline{M} \equiv [M(D^{0})+M(\bar{D}^{0})]/2$ closest to the nominal $D$ mass is selected.

\subsection{\boldmath Selection of partially reconstructed tags} \label{sec:partialselection} 
Partially reconstructed tags collectively refer to the tag modes where there is one particle in the final state, either a $K_{\rm L}^{0}$ meson or a neutrino, which is not reconstructed. Modes with more than one missing particle in the final state are not considered in this analysis. Due to the presence of a missing particle in the final state, these tag modes can be reconstructed only as DTs so that four-momentum conservation can be exploited in the reconstruction. 

\begin{figure*}[ht!] \centering
\includegraphics[width=6.0in,height=6.0in,keepaspectratio]{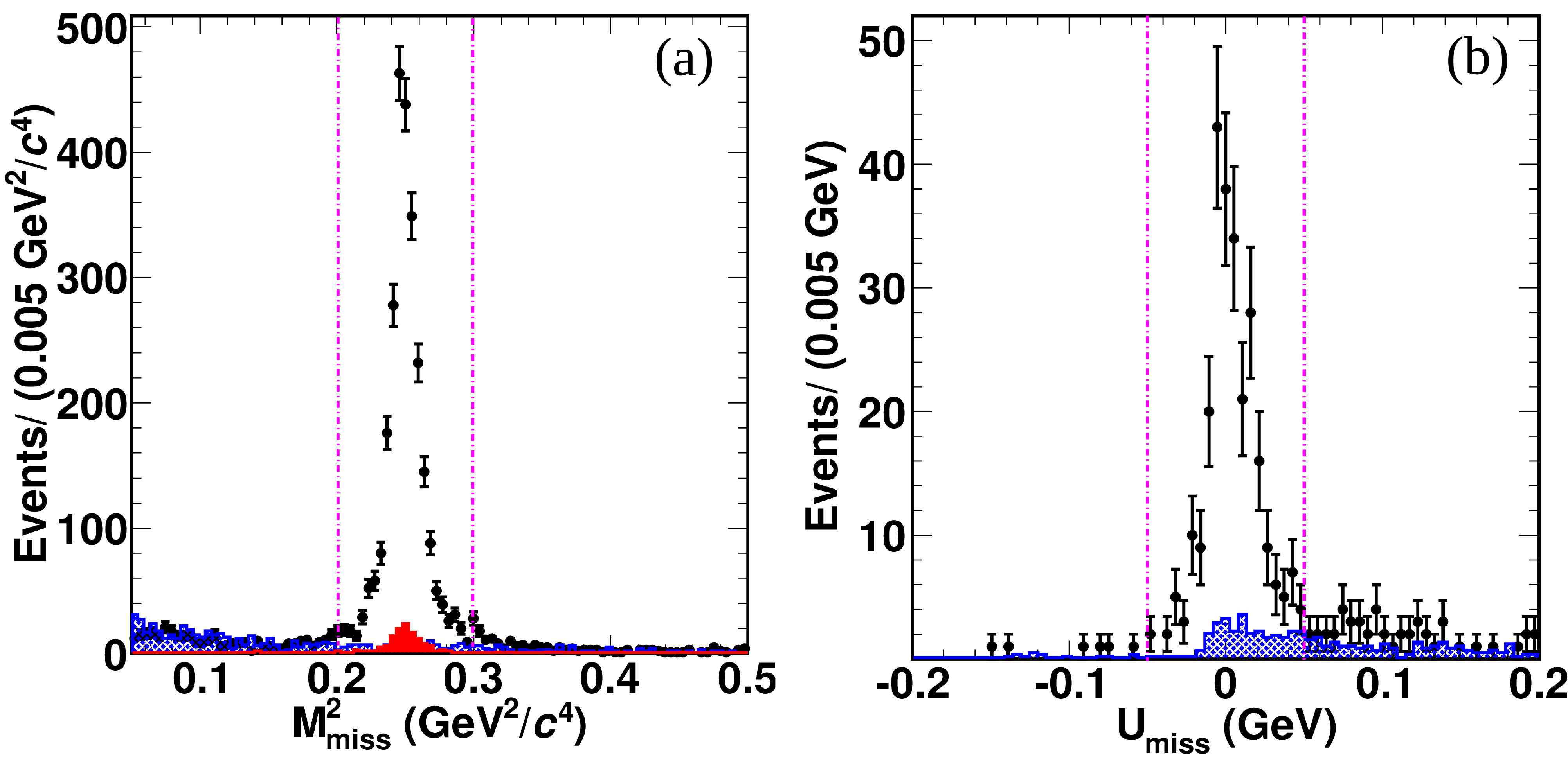}
\caption{(a) $M_{\rm miss}^{2}$ distribution for $\bar{D}^0\to K^{0}_{\rm L}K^{+}K^{-}$ candidates reconstructed against the flavor-tags $D^0\to K^{-}\pi^+$ and $D^0\to K^{-}\pi^{+}\pi^0$. The points with error bars are the data, the red histogram denotes the peaking background due to $\bar{D}^0\to K^{0}_{\rm S}K^{+}K^{-}$ events from the inclusive MC sample, the blue-shaded histogram shows the combinatorial backgrounds from inclusive MC samples, and the magenta vertical lines indicates the signal region. (b) $U_{\rm miss}$ distribution for events in which $\bar{D}^0\to K^{0}_{\rm S}K^{+}K^{-}$ candidates are reconstructed against the $D^0\to K^{-}e^{+}\nu_{e}$ tag. The black points with error bars are data, the blue-shaded histogram shows the backgrounds estimated from the inclusive MC sample and the magenta vertical line shows the signal region.}
\label{fig:mmissandumiss}
\end{figure*}

Selections of partially reconstructed tag modes proceed as follows. The opposite-side $D$ candidate is reconstructed as a ST candidate using the criteria given in the  Sec.~\ref{sec:fullselection}. All the particles except the missing particle in the final state are reconstructed from the unused tracks and showers that satisfy the selection criteria already described. The presence of an unreconstructed $K^0_{\rm L}$ is inferred from the missing-mass distribution, calculated from the missing energy, $E_{\rm miss}$, and the missing momentum, $\mathbf{p}_{\rm miss}$, in the center-of-mass frame as
\begin{equation}\label{eq:klemisseqn}
M^{2}_{\rm miss} \equiv \frac{E^{2}_{\rm miss}}{c^4} - \frac{\left|\mathbf{p}_{\rm miss}\right|^{2}}{c^2},
\end{equation}
which peaks at $m_{K^0}^2$ for signal, where $m_{K^0}$ is the nominal mass of the neutral kaon given in Ref.~\cite{pdg}. 
The presence of a neutrino is inferred using the variable
\begin{equation}\label{eq:umisseqn}
U_{\rm miss} \equiv E_{\rm miss} - \left|\mathbf{p}_{\rm miss}\right|c \;,
\end{equation}
which peaks at zero for signal. Again we take advantage of resonant production and the knowledge of beam energy to determine $E_{\rm miss}$ and $\mathbf{p}_{\rm miss}$. Figure~\ref{fig:mmissandumiss} shows example distributions of $M^{2}_{\rm miss}$ and $U_{\rm miss}$.
Reconstruction using the missing-mass technique inevitably results in a higher level of background than the full-reconstruction method. To reduce the background further, we do not consider events that have more charged tracks or neutral particles than required in the final state. The angle, $\alpha$, between the $\mathbf{p}_{\rm miss}$ and the nearest unassigned shower is calculated. All the events with $\cos\alpha$ $>$ 0.98 are retained. For the events with $\cos\alpha$ $<$ 0.98 mode-dependent criteria are applied on the energy of the unassigned shower. Even though we reject events with additional neutral particles in the final state, there is significant background in the modes with neutral particles in the final states, arising from the final states having additional neutral particles that are not reconstructed. For example, in the case of $D\to K^{0}_{\rm L}\pi^{0}$ decays there are backgrounds from $K^{0}_{\rm L}\pi^{0}\pi^{0}$ where one $\pi^{0}$ meson is not reconstructed, so the event passes all our selection criteria. These backgrounds can be further reduced by applying criteria on the momentum spectrum of reconstructed $\pi^{0}$ or $\eta$ candidates wherever applicable. The values of these criteria are selected based on optimization studies  that use the inclusive MC samples. This optimization maximizes the figure-of-merit defined as $S/\sqrt{S+B}$, where $S~(B)$ are the number of signal (background) events in the signal region retained by the selection; the signal region for the optimization is the interval $0.2<M_{\rm miss}^2 <0.3~\mathrm{GeV}^2/c^4$. The values of the shower energy and $\pi^0$ $(\eta)$ momentum criteria are varied, and the value that maximizes the figure-of-merit is chosen.

\section{\boldmath Estimation of ST and DT yields}\label{sec:yieldestimation} 
In Secs.~\ref{sec:styield} and \ref{sec:dtyield} we will describe the methods of estimating ST and DT yields, respectively. Note that DT yields are only required bin-by-bin, not integrated over the Dalitz plot.

\subsection{\boldmath ST yields}\label{sec:styield}
 ST yields of fully reconstructed tag modes are determined from maximum likelihood fits to the $M_{\rm{BC}}$ distribution. Our probability density function (PDF) is a sum of the signal shape derived from the signal MC sample convolved with a Gaussian function to account for any difference in resolution between data and MC simulation, and an ARGUS function~\cite{argus} to model the background. The threshold of the ARGUS function is fixed at $M_{BC}=1.8865~$GeV$/c^{2}$, which corresponds to the kinematic limit of $D^{0}$ production at the $\psi(3770)$. The peaking background is modelled by the shapes and yields obtained from the inclusive MC sample; this assumption is considered as a source of systematic uncertainty. The flavor-tag modes $D^0\to K^{-}\pi^{+}$ and $D^0\to K^{-}\pi^{+}\pi^{0}$ have a peaking background of approximately 0.2\% from DCS decays. The dominant peaking background to the decays $D\to K^{0}_{\rm S}\pi^{0}$ and $D\to K^{0}_{\rm S}\pi^{0}\pi^{0}$ is from $D\to \pi^{+}\pi^{-}\pi^{0}$ (0.5\%) and $D\to \pi^{+}\pi^{-}\pi^{0}\pi^0$ (7\%) decays, respectively.  The $M_{\rm BC}$ distribution is fit over the range (1.83,~1.88)~GeV$/c^{2}$. The ST yields are obtained by integrating the $M_{\rm BC}$ distribution in the range (1.86,~1.87)~GeV$/c^{2}$. In order to eliminate the small effect of $D^{0}\bar{D}^{0}$ mixing, the measured ST yields of $CP$ modes are multiplied by a correction factor of $1/(1-\eta_{\pm}y_{D})$, where $\eta_{\pm}$ is the $CP$ eigenvalue of the mode and $y_{D}$ is the charm-mixing parameter taken from Ref.~\cite{pdg}.

\begin{figure*}
\includegraphics[width=\linewidth]{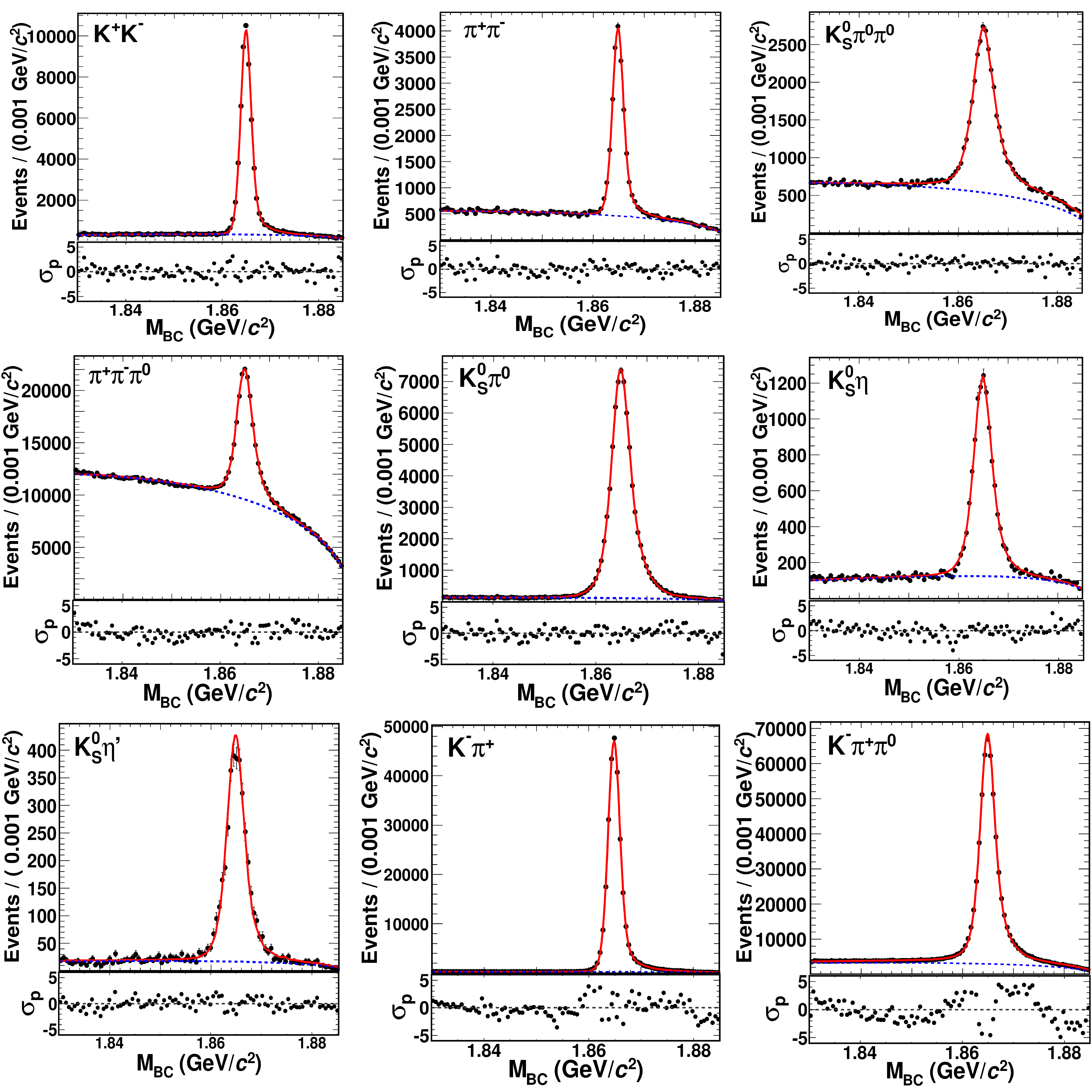}
\caption{Fits to the $M_{\rm BC}$ distributions of ST decay modes. The points with error bars are data, the red curve is the total fit result and the blue dashed curve is the background component. Beneath each distribution the pull ($\sigma_{\rm p}$) between the data and the fit is shown. The significant pulls observed in the flavor-tag modes $D^0\to K^-\pi^+$ and $D^0\to K^-\pi^+\pi^0$ are a consequence of the large sample size but studies of MC samples indicate that there is no significant bias on the ST yield introduced as a result.}
\label{fig:stfits}
\end{figure*}

The ST yield, $S_{\rm ST}$, of a partially reconstructed tag is calculated using the relation
\begin{equation}\label{eq:klsingletagyield}
S_{\rm ST}= 2N_{D^{0}\bar{D}^{0}}\mathcal{B}_{\rm ST},
\end{equation}
where $N_{D^{0}\bar{D}^{0}}$ is the number of $D^{0}\bar{D}^{0}$ pairs in the BESIII data sample \cite{besNDDbar} and $\mathcal{B}_{\rm ST}$ is the branching fraction of the tag mode, which is taken from Ref.~\cite{pdg} where available. The branching fractions of all $D\to K^{0}_{\rm L}X$ modes except $D\to K^{0}_{\rm L}\pi^{0}$ are not available in Ref.~\cite{pdg}, hence we assume the branching fractions of these modes to be the same as for the corresponding $D\to K^{0}_{\rm S}X$ modes.
We note that this reasoning is not strictly valid, as the interference between Cabibbo-favored  (CF) ($D^{0}\rightarrow \bar{K^{0}}X$) and DCS transitions ($D^{0}\rightarrow K^{0}X$) can lead to a difference in the decay rates for $D\rightarrow K^{0}_{\rm L}X$ and $D\rightarrow K^{0}_{\rm S}X$. However, this difference is expected to be less than 10\% ~\cite{bigi}, which is considered as a systematic uncertainty; the difference will barely affect our final results, as the ST yields are used only for yield normalisation, as given in Eqs.~(\ref{eq:Mieqn}) and~(\ref{eq:Miprimeeqn}). The ST yields calculated using Eq.~(\ref{eq:klsingletagyield}) have larger uncertainties compared to the fully reconstructed tags, largely due to the uncertainty of the assumed values of $\mathcal{B}_{\rm ST}$. The ST $M_{\rm BC}$ fits are shown in Fig.~\ref{fig:stfits} and the yields are given in Table~\ref{tab:yieldsandefficiencies}. The effect on the final measurement due to the uncertainty in the measured values of the ST yields is treated as a systematic uncertainty. The ST yield uncertainty includes systematic uncertainties related to the fit procedure.

\subsection{\boldmath DT yield}\label{sec:dtyield}
\begin{figure}[ht!] \centering
\includegraphics[width=3in]{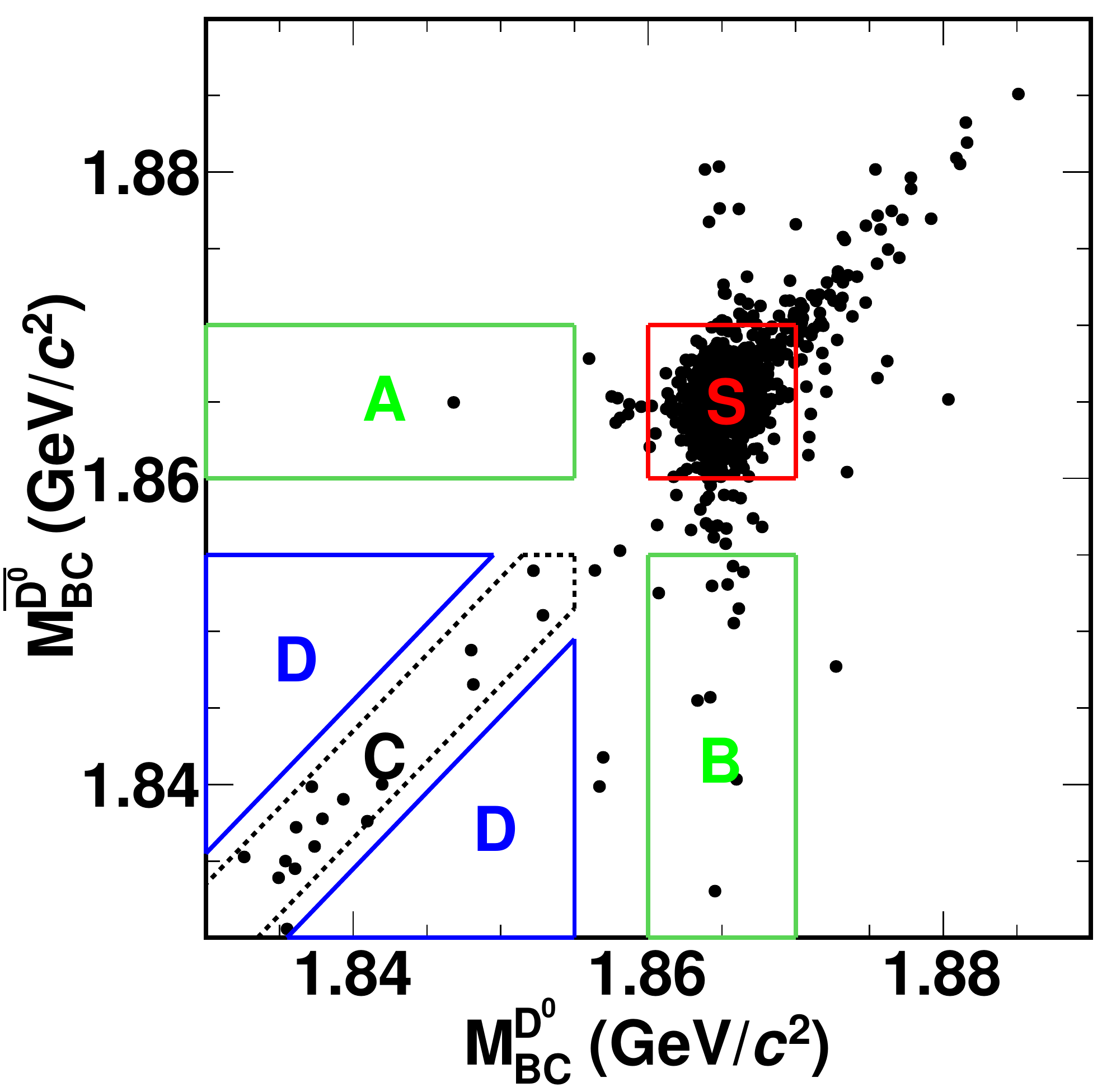}
\caption{Distribution of events across $(M_{\rm BC}^{D^{0}},M_{\rm BC}^{\bar{D}^{0}})$ plane for $K_{\rm S}^{0}K^{+}K^{-}$ reconstructed against flavor-tags. The signal region is denoted as S, while A, B, C and D are the various sideband regions.}
\label{fig:Dtmbcdist}
\end{figure}

For fully reconstructed DT modes we follow a sideband-estimation method developed by the CLEO Collaboration~\cite{cleo} to determine the DT yield. The sidebands are defined on the two-dimensional $(M_{\rm BC}^{D^{0}},~M_{\rm BC}^{\bar{D}^{0}})$ plane as shown in Fig.~\ref{fig:Dtmbcdist}. Here, the $M_{\rm BC}^{D^{0}}$ ($M_{\rm BC}^{\bar{D}^{0}}$) refers to the $M_{\rm BC}$ distribution of signal (tag) side. In Fig.~\ref{fig:Dtmbcdist}, S refers to the signal region, sideband A (B) contains events which are from misreconstructed tag (signal) decays, sideband C consists of continuum events and sideband D consists of events that are purely combinatoric. The amount of combinatorial (non-peaking) backgrounds in the signal region is estimated from the events in the sideband regions. Thus the total DT yield, $N_{\rm DT}$, of $K^{0}_{\rm S}K^{+}K^{-}$ is estimated as 
\begin{eqnarray}
&&N_{\rm DT}  =  N_{\rm S} - N_{\rm P} \nonumber \\ &&- \left(\frac{a_{\rm S}}{a_{\rm D}} N_{\rm D} + \sum_{i = \rm A, B, C}\frac{ a_{\rm S}}{a_{i}}\left(N_i - \frac{ a_{\rm S}}{a_{i}} N_{\rm D} \right)\right) \;,
\end{eqnarray}
where $a_i$ is the area of the corresponding region $i=$ A, B, C, D or S, $N_i$ refers to the yields in the sideband region, $N_{\rm S}$ is the yield in the signal region before background correction (uncorrected yield) and $N_{\rm P}$ is the peaking-background yield estimated from the MC simulation (see Sec.~\ref{sec:binbkgestimation}).

In the case of partially reconstructed tag modes we follow a similar sideband-estimation method as in Ref.~\cite{chris_thesis}. Here three regions are defined on the $M_{\rm miss}^{2}$ or $U_{\rm miss}$ distributions: low sideband (L), signal region (S) and high sideband (H). The total yield is estimated as
\begin{equation}
Y_{\rm S} = \frac{(N_{\rm S} - N_{\rm S}^{\rm P}) - \delta(N_{\rm L} - N_{\rm L}^{\rm P})  - \gamma(N_{\rm H} - N_{\rm H}^{\rm P})}{1 - \delta\alpha - \gamma\beta}\;, 
\end{equation} 
where $N_{\rm S}$, $N_{\rm L}$ and $N_{\rm H}$ are the uncorrected yields in the signal and sideband regions, $N^{\rm P}_{i}$ refers to the peaking background in the $i$th region, $\delta$ and $\gamma$ refer to the ratio of combinatorial backgrounds in the signal region to that in the L and H sideband regions, respectively, and $\alpha$ and $\beta$ refer to the ratio of signal in region S to that in the regions L and H, respectively. The values of $\alpha$, $\beta$, $\delta$ and $\gamma$ are derived from MC samples. Here the definitions of sidebands are mode dependent. We follow the same optimization procedure described in Sec.~\ref{sec:partialselection} to define the signal regions. The peaking backgrounds are estimated from MC samples as described in Sec.~\ref{sec:binbkgestimation}.

\begin{table*}[t]
 \begin{center}
 \caption{Single-tag (ST) and $D^{0}\to K^{0}_{\rm S,L}K^{+}K^{-}$ double-tag (DT) yields and efficiencies. The DT yields are the observed number of events in the signal region prior to background 
and efficiency corrections. The ST yields are background subtracted because they are the result of fits to the $M_{\rm BC}$ distributions. }\label{tab:yieldsandefficiencies}
\setlength{\tabcolsep}{6.5pt}
  \begin{tabular}{l c c | c c c c} 
  \hline
  \hline
  Mode & \multicolumn{2}{c|}{ST} & \multicolumn{3}{c}{DT} \\
       &  $N_{\rm ST}$  & $\epsilon_{\rm ST}$(\%)   & $N^{K^{0}_{\rm S}K^{+}K^{-}}_{\rm DT}$ & $N^{K^{0}_{\rm L}K^{+}K^{-}}_{\rm DT}$ & $\epsilon_{\rm DT}^{K^{0}_{\rm S}K^{+}K^{-}}$(\%) & $\epsilon_{\rm DT}^{K^{0}_{\rm L}K^{+}K^{-}}$(\%) \\ 
\hline
Flavor-tags & & & & & &\\
$K^{-}\pi^{+}$ & $524307\pm 742$ & $63.31 \pm 0.06$ & 323 & 743 & 12.43 $\pm$ 0.07 & 15.85 $\pm$ 0.08 \\
$K^{-}\pi^{+}\pi^{0}$ & $\phantom{1}995683 \pm 1117 $ & $31.70 \pm 0.03$ & 596 & 1769 & \phantom{1}5.86 $\pm$ 0.05 & \phantom{1}7.94 $\pm$ 0.06 \\
$K^{-}e^{+}\nu_e$ & $\phantom{01}752387 \pm 12795$ &  & 263 & & \phantom{1}3.23 $\pm$ 0.04 & \\ \hline
$CP$-even tags & & & & & &\\
$K^{+}K^{-}$ & $\phantom{1}53481 \pm 247$ & $61.02 \pm 0.11$ & 42 & 112 & 12.07 $\pm$ 0.07 & 15.52 $\pm$ 0.08 \\
$\pi^{+}\pi^{-}$ & $\phantom{1}19339 \pm 163$ & $64.52 \pm 0.11$ & 10 & 31 & 12.16 $\pm$ 0.07 & 15.70 $\pm$ 0.08 \\
$K^{0}_{\rm S}\pi^{0}\pi^{0}$ & $\phantom{1}19882 \pm 233$& $14.86 \pm 0.08$ & 7 & 45 & \phantom{1}2.49 $\pm$ 0.04 & \phantom{1}3.79 $\pm$ 0.04\\
$\pi^{+}\pi^{-}\pi^{0}$ &\phantom{1}99981 $\pm$ 618 & $37.65 \pm 0.11$ & 51 & 254 & \phantom{1}6.79 $\pm$ 0.06 & \phantom{1}9.54 $\pm$ 0.07 \\
$K^{0}_{\rm L}\pi^{0}$ & $\phantom{01}209445 \pm 14796$ &  & 90 & & \phantom{1}8.88 $\pm$ 0.06 &   \\
$K^{0}_{\rm L}\eta(\gamma\gamma)$ & $\phantom{01}40009 \pm 2543$ &  & 19 &  & \phantom{1}6.60 $\pm$ 0.06 & \\
$K^{0}_{\rm L}\omega$ & $\phantom{01}207376 \pm 11498$ &  & 44 &  & \phantom{1}3.42 $\pm$ 0.04  &\\
$K^{0}_{\rm L}\eta^{\prime}(\pi^{+}\pi^{-}\eta)$ & $\phantom{01}33683 \pm 1909$ &  & 7 &  & \phantom{1}3.23 $\pm$ 0.04 & \\ 
\hline
$CP$-odd tags & & & & & &\\
$K^{0}_{\rm S}\pi^{0}$ & $65072 \pm 281$ & $36.92 \pm 0.11$ & 39 & 89  & \phantom{1}6.75 $\pm$ 0.06 & 9.33 $\pm$ 0.07\\
$K^{0}_{\rm S}\eta(\gamma\gamma)$ & $9524 \pm 134$ & $32.94 \pm 0.11$ & 9 & 10 & \phantom{1}6.05 $\pm$ 0.05 & 9.05 $\pm$ 0.06\\
$K^{0}_{\rm S}\omega$ & $\phantom{0}19262 \pm 157$& $12.14 \pm 0.07$ & 16 &  27 & \phantom{1}2.20 $\pm$ 0.03 & 3.42 $\pm$ 0.04 \\
$K^{0}_{\rm S}\eta^{\prime}(\pi^{+}\pi^{-}\eta)$ & $\phantom{0}3301 \pm 62$ & 12.46 $\pm$ 0.07 & 2 & 5 & \phantom{1}2.20 $\pm$ 0.03 & 3.46 $\pm$ 0.04 \\
\hline
Mixed $CP$ tags & & & & & &\\
$K^{0}_{\rm S}\pi^{+}\pi^{-}$ & &  & 78 & 265  & \phantom{1}6.35 $\pm$ 0.05 & 8.32 $\pm$ 0.06 \\
$K^{0}_{\rm L}\pi^{+}\pi^{-}$ & & & 282 &  & \phantom{1}9.56 $\pm$ 0.07 & \\
$K^{0}_{\rm S}K^{+}K^{-}$ & \phantom{0}12949 $\pm$ 119 & $18.35 \pm 0.09$ & 4 & 19  & \phantom{1}2.99 $\pm$ 0.04 & 3.40 $\pm$ 0.04\\ \hline
\hline
  \end{tabular}
 \end{center}
\end{table*}

\section{\boldmath $D\to K^{0}_{\rm S, L}K^{+}K^{-}$ Dalitz plots}\label{sec:dalitzplot}
\begin{figure*}[ht!] \centering
\includegraphics[width=5in,height=5in,keepaspectratio]{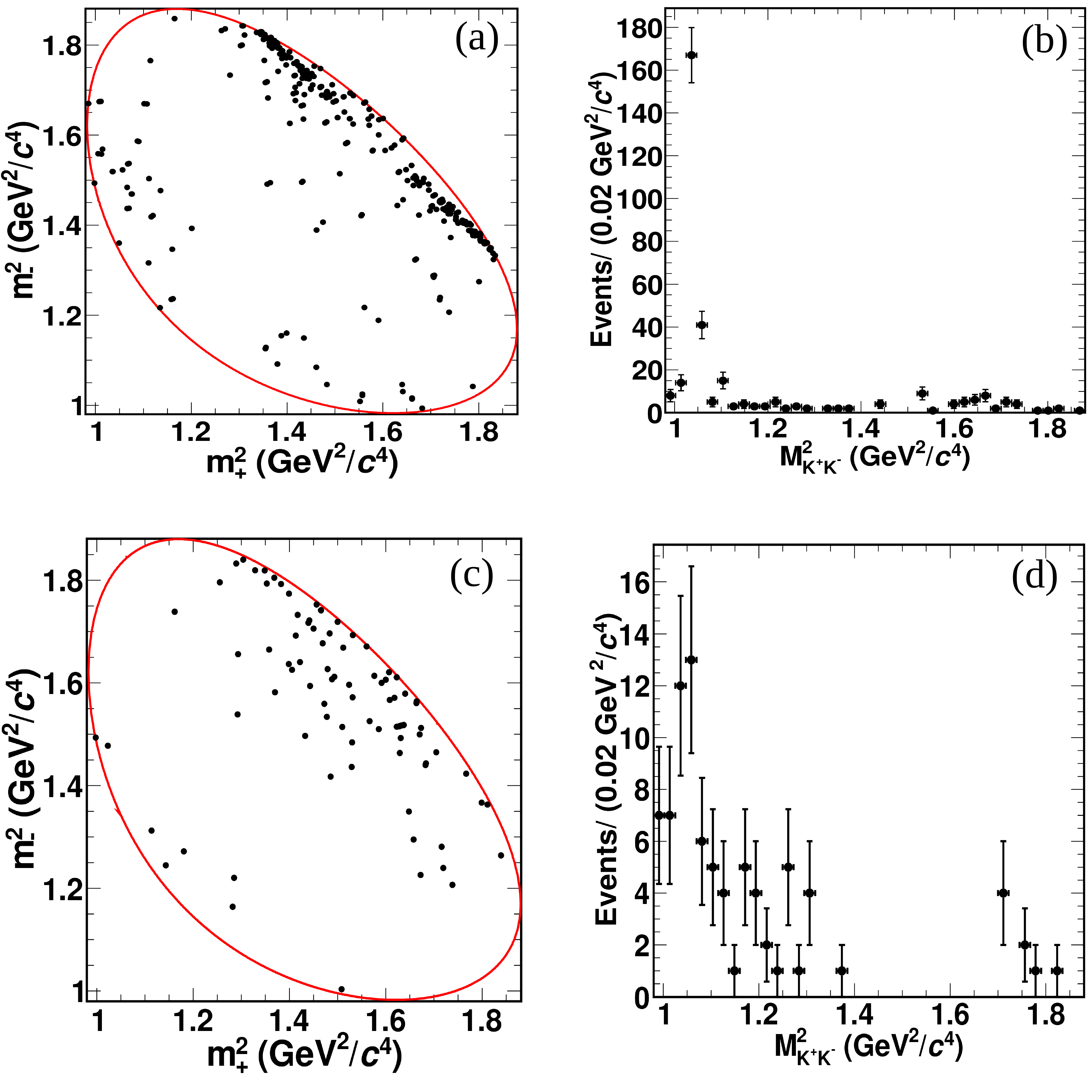}
\caption{(a) Dalitz plot and (b) $M^{2}_{K^{+}K^{-}}$ distributions for $D\to K^{0}_{\rm S}K^{+}K^{-}$ reconstructed against $CP$-even final states. (c) Dalitz plot and (d) $M^{2}_{K^{+}K^{-}}$ distributions for $D\to K^{0}_{\rm S}K^{+}K^{-}$ reconstructed against $CP$-odd final states. } 
\label{fig:K0SKK_Dalitz_CPeven}
\end{figure*}

\begin{figure*}[ht!] \centering
\includegraphics[width=5in,height=5in,keepaspectratio]{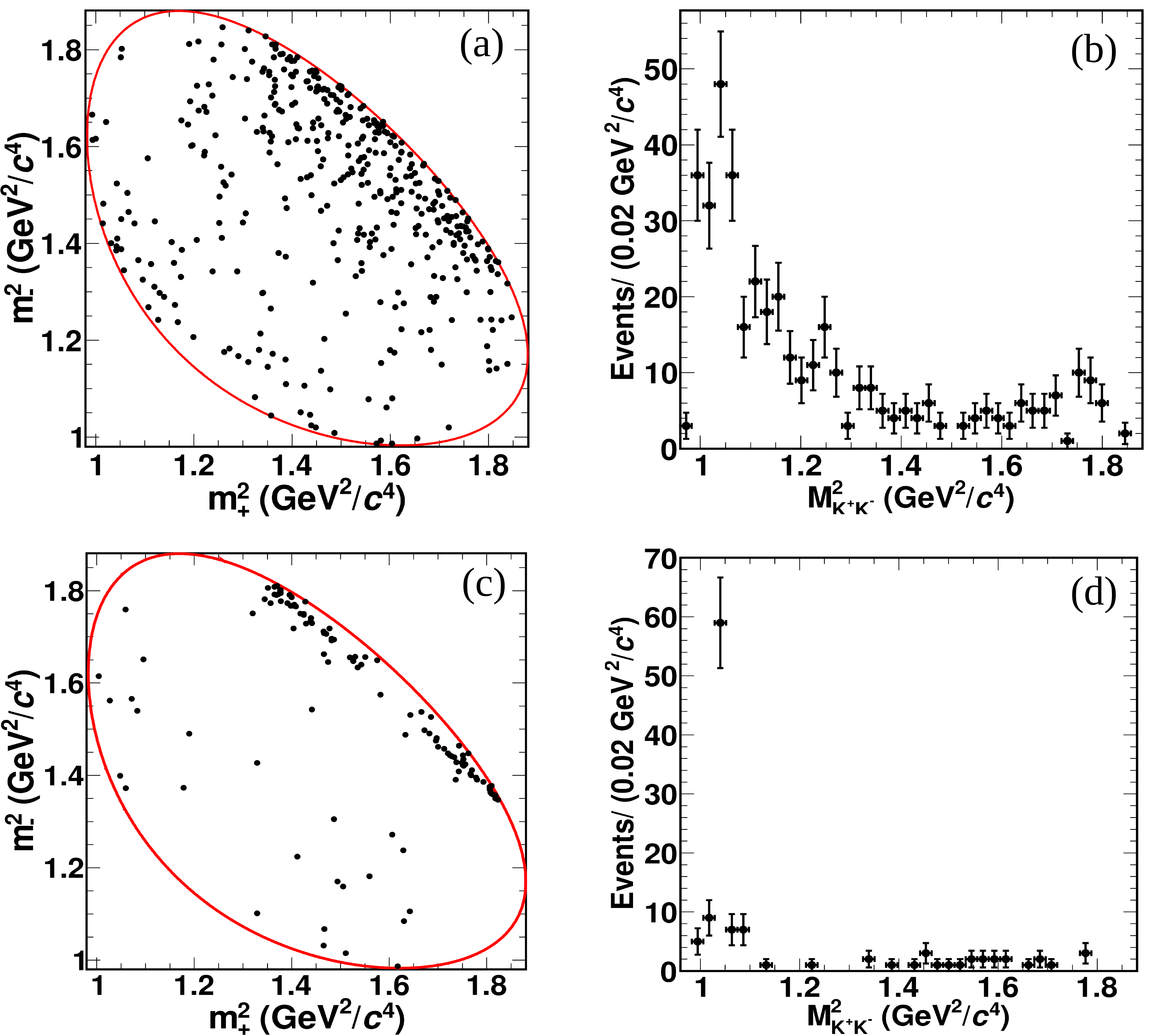}
\caption{(a) Dalitz plot and (b) $M^{2}_{K^{+}K^{-}}$ distributions for $D\to K^{0}_{\rm L}K^{+}K^{-}$ reconstructed against $CP$-even final states. (c) Dalitz plot and (d) $M^{2}_{K^{+}K^{-}}$ distributions for $D\to K^{0}_{\rm L}K^{+}K^{-}$ reconstructed against $CP$-odd final states. }
\label{fig:K0SKK_Dalitz_CPodd}
\end{figure*}

\begin{figure*}[ht!] \centering
\includegraphics[width=5in,height=5in,keepaspectratio]{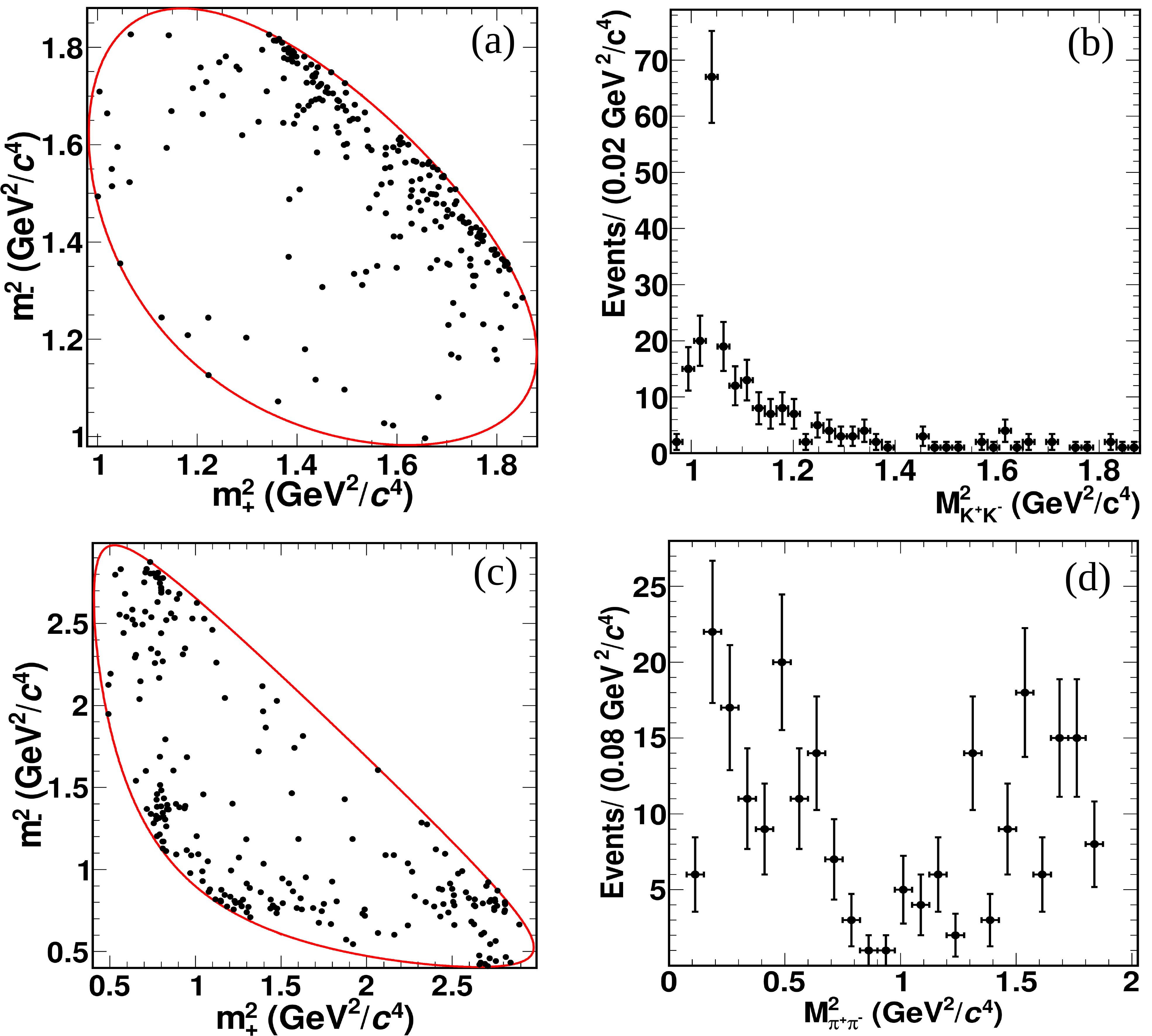}
\caption{(a) Dalitz plot and (b) $M^{2}_{K^{+}K^{-}}$ distributions for $D\to K^{0}_{\rm L}K^{+}K^{-}$ reconstructed against $\bar{D}\to K^{0}_{\rm S}\pi^{+}\pi^{-}$ final states. (c) Dalitz plot and (d) $M^{2}_{\pi^{+}\pi^{-}}$ distributions for $D\to K^{0}_{\rm S}\pi^{+}\pi^{-}$ decay in the same events. }
\label{fig:KLSKK_Dalitz_Selfconjugate}
\end{figure*}
 
In this section we discuss the Dalitz plot distributions of events when $D\to K^{0}_{\rm S, L}K^{+}K^{-}$ candidates are tagged with pure $CP$ eigenstates and mixed $CP$ states; we highlight the important differences. 

In order to improve the resolution on the Dalitz plot variables $(m_+^2,m_-^2)$, a kinematic fit is performed for $D \to K^{0}_{\rm S,L}h^{+}h^{-}$ candidates. For $D\to K^{0}_{\rm S}h^{+}h^{-}$ tags, the two pions from the $K^{0}_{\rm S}$ candidate obtained after the secondary vertex fit are combined with the $h^{+}$ and $h^{-}$ into a common fit to the nominal mass of the $D^{0}$ meson taken from Ref.~\cite{pdg}. In the case of a $D\to K^{0}_{\rm L}h^{+}h^{-}$ candidate, a missing particle is created using the position of an EMC shower associated with the $K_{\rm L}^0$ candidate. The mass of this object is set to the nominal mass of the $K^{0}_{\rm L}$ meson taken from Ref.~\cite{pdg}; it is combined with $h^{+}h^{-}$ tracks and fit to the nominal mass of the $D^{0}$ meson. A 35 to 40\% (30 to 35\%) improvement in the $m_{\pm}^2$ resolution across the Dalitz plot is achieved for $D\to K_{\rm S}^{0}K^{+}K^{-}$  ($D \to K_{\rm L}^{0}K^{+}K^{-}$) candidates after the kinematic fit. The resolutions are quantified using the signal MC samples.
Events that fail  the kinematic fit are rejected.
The improved values of $(m_+^2,m_-^2)$ are used to define the position of the event within the Dalitz plot and assign its bin index. 

The Dalitz plot distribution of the $D\to K^{0}_{\rm S}K^{+}K^{-}$ candidates reconstructed against $CP$-even tag modes and their corresponding $M^{2}_{K^{+}K^{-}}$ projections are given in Fig.~\ref{fig:K0SKK_Dalitz_CPeven}. The presence of a significant peak around $M^{2}_{K^{+}K^{-}}\sim~1.04~\mathrm{GeV}^2/c^{4}$ is due to the decay $D^0 \to K^{0}_{\rm S}\phi,~\phi\to K^{+}K^{-}$. These events are distributed along the diagonal boundary of the Dalitz plot.
As $D^{0}\to K^{0}_{\rm S}\phi$ constitutes a large fraction of the total $D^{0}\to K^{0}_{\rm S}K^{+}K^{-}$ decay width \cite{pdg}, a higher population of events is seen in the region enclosing the $\phi$ resonance.
%
A similar peak is absent in the $M^{2}_{K^{+}K^{-}}$ distribution of $D \to K^{0}_{\rm S}K^{+}K^{-}$ candidates reconstructed against $CP$-odd tag modes shown in Figs.~\ref{fig:K0SKK_Dalitz_CPeven}(c) and (d). This is a consequence of the quantum-correlation in data. Since each $D$ meson is of opposite $CP$ eigenvalue,  the $K^{0}_{\rm S}K^{+}K^{-}$ candidates reconstructed against $CP$-odd tags should decay through $CP$-even intermediate states. Hence it cannot decay through the $D\to K^{0}_{\rm S}\phi$ state. The dominant $CP$-even intermediate state is the $D\to K^{0}_{\rm S}a(980)^0$ decay. The distribution of events in the Dalitz plot is observed to be flatter than in the case of $K^{0}_{\rm S}K^{+}K^{-}$ tagged against a $CP$-even state. Since $K_{\rm S}^{0}$ and $K_{\rm L}^{0}$ have opposite $CP$ eigenvalues, the entire scenario is reversed in the case of $D\to K_{\rm L}^{0}K^{+}K^{-}$ decays as shown in Fig.~\ref{fig:K0SKK_Dalitz_CPodd}. The Dalitz plot distribution of $D\to K_{\rm L}^{0}K^{+}K^{-}$ candidates against $CP$-even modes resembles that of   $D \to K_{\rm S}^{0}K^{+}K^{-}$ candidates against $CP$-odd modes and vice versa. 

The Dalitz plot distribution of $D\to K_{\rm L}^{0}K^{+}K^{-}$ candidates against the self-conjugate mode $D\to K_{\rm S}^{0}\pi^{+}\pi^{-}$ is given separately for signal and tag sides in Fig.~\ref{fig:KLSKK_Dalitz_Selfconjugate}. The Dalitz plot of $D\to K_{\rm S}^{0}\pi^{+}\pi^{-}$ tags is consistent with that presented in Ref.~\cite{lei}.  In  Fig.~\ref{fig:KLSKK_Dalitz_Selfconjugate}(d), the enhancement above $M_{\pi^{+}\pi^{-}}^{2}\sim$ 1.3 GeV$^{2}/c^{4}$ corresponds to $D\to K^{*}(892)^{\pm}\pi^{\mp}$ decays, whereas the peak around $M_{\pi^{+}\pi^{-}}^{2}\sim$ 0.6 GeV$^{2}/c^{4}$ corresponds to 
$D\to K^{0}_{\rm S}\rho^{0}$ decays. The $D\to K^{*}(892)^{\pm}\pi^{\mp}$ decays can be seen as two bands that are parallel to the vertical and horizontal axes of the Dalitz plane. The decay $D\to K^{0}_{\rm S}\rho^{0}$ lies close and parallel to the diagonal boundary. Since the $D\to K_{\rm L}^{0}K^{+}K^{-}$ decays reconstructed against $D\to K_{\rm S}^{0}\pi^{+}\pi^{-}$ decays are not in a $CP$ eigenstate, the Dalitz plot distribution is a combination of both the $CP$-even and $CP$-odd tagged $K_{\rm L}^{0}K^{+}K^{-}$ Dalitz plots. The Dalitz plot structure of $D\to K^{0}_{\rm S}K^{+}K^{-}$ reconstructed against $D\to K^{0}_{\rm S,L}\pi^{+}\pi^{-}$ has similar features to those shown in Fig.~\ref{fig:KLSKK_Dalitz_Selfconjugate}.

\subsection{\boldmath Dalitz plot binning, bin yield estimation and corrections}\label{sec:binyieldandcorrection}
In this section we describe our method of binning the Dalitz plots and calculating the bin yields and efficiencies. The procedures for correcting the bin migration and DCS correction for flavor-tag yields are also explained.  

The binning prescription followed in our analysis is described in Sec.~\ref{sec:formalism}. The entire $D^0\to K_{\rm S,L}^{0}K^{+}K^{-}$ Dalitz plot is divided into $\mathcal{N} = 2,~\mathcal{N} = 3$ and $\mathcal{N} =4$ equal-$\Delta\delta_{D}$ bins. In the case of $D\to K^{0}_{\rm S,L}\pi^{+}\pi^{-}$ tag modes, the entire Dalitz plot is divided into $\mathcal{N} = 8$ equal-$\Delta\delta_{D}$ bins identical to those defined in Ref.~\cite{lei}. The uncorrected bin yields are obtained by counting the number of events in each bin. The bin yield $M_{i}^{\pm}$ (see Eq.~(\ref{eq:Mieqn})) of $D
\to K_{\rm S}^{0}K^{+}K^{-}$ reconstructed against $CP$ tags and the flavor-tag yield, $K_{i}$ (see Eq.~(\ref{eqn:relationkiandti})) are calculated separately for each mode. The events in the $i$th bin of the $D\to K^{0}_{\rm S}K^{+}K^{-}$ Dalitz plot and the $j$th bin of the $D\to K_{\rm S}^{0}h^{+}h^{-}$ Dalitz plot are counted to obtain $M_{ij}$ (see Eq.~(\ref{eq:mijeqn})). A similar procedure is followed to obtain the yields $K_{i}^{\prime}$, $M_{i}^{\pm\prime}$ and $M_{ij}^{\prime}$ (see Eqs.~(\ref{eq:Miprimeeqn}) and~(\ref{eq:Mijprimeeqn})) for the $D\to K^{0}_{\rm L}K^{+}K^{-}$ decay. The flavor-tag yield for the $D^{0}\to K^{0}_{\rm S,L}\pi^{+}\pi^{-}$ mode is taken from Ref.~\cite{lei}. The yields of $D
\to K^{0}_{\rm S}K^{+}K^{-}$ decays reconstructed against $CP$ tags are quite low. The inclusion of the $D\to \pi^{+}\pi^{-}\pi^{0}$ tag mode results in an approximately 50\% increase in the $CP$-even tag yield. The uncorrected yields of $D\to K^{0}_{\rm S}K^{+}K^{-}$ decays reconstructed against $CP$ tags, along with their efficiencies, are given in Table~\ref{tab:yieldsandefficiencies}. 

Due to the finite $(m_+^2,m_-^2)$ resolution, events migrate between bins. Often these migrations are asymmetric between the bins because of the differing event densities in each bin. We correct for this using an unfolding method based on correction factors derived from the signal MC samples. For $D\to K_{\rm S,L}^{0}K^{+}K^{-}$ decays reconstructed against $CP$ and flavor tags, we define a $2\mathcal{N}\times 2\mathcal{N}$ migration matrix {\bf U} as
\begin{equation}\label{eq:migrationmatrixdefinition}
U_{i,j} \equiv \frac{m_{ji}}{\sum_{k = -\mathcal{N},k \neq 0}^{\mathcal{N}}m_{jk}},
\end{equation}
where $m_{ji}$ are the events generated in the $j$th bin and reconstructed in the $i$th bin. The vector of migration-corrected data yields $\mathrm{\textbf{N}}$ and the vector of reconstructed yields in the signal region $\mathrm{\textbf{N}_{\rm S}}$ are related by
  \begin{equation}
  {\mathrm{\textbf{N}} = \mathrm{\textbf{U}^{-1}\textbf{N}_{\textbf{\rm S}}}}.
   \end{equation}
In the case of $D\to K_{\rm S,L}^{0}K^{+}K^{-}$ reconstructed against the $D\to K_{\rm S,L}^{0}h^{+}h^{-}$ tags, the correlation between the bins on the signal and tag sides needs to be taken into account. Hence the total migration matrix is a tensor (Kronecker) product of signal- and tag-migration matrices. For a given number of signal bins $\mathcal{N}$, the dimension of the migration matrix for $K^{0}_{\rm S,L}K^{+}K^{-}$ against $K^{0}_{\rm S,L}K^{+}K^{-}$ is 4$\mathcal{N}^{2}$ $\times$ 4$\mathcal{N}^{2}$ and for $K^{0}_{\rm S,L}K^{+}K^{-}$ against $K^{0}_{\rm S,L}\pi^{+}\pi^{-}$ it is 32$\mathcal{N} \times 32\mathcal{N}$. The uncertainties in the matrix elements due to the finite size of the signal MC sample are considered as a source of systematic uncertainty. An example of the migration matrix for $D\to K^{0}_{\rm  S}K^{+}K^{-}$ candidates reconstructed against the $D\to K^{+}K^{-}$ tag mode is given in Table~\ref{tab:migmatrix}. Typically the rate of migration out of bin 1, which contains the narrow $\phi$ resonance, is about 3\% for $D\to K^{0}_{\rm  S}K^{+}K^{-}$ decays and about 5\% for $D\to K^{0}_{\rm L}K^{+}K^{-}$ decays. The rate of migration into bin 1 is significantly smaller due to the broader structures that occupy the remainder of the Dalitz plot away from the $\phi$ resonance. Throughout this unfolding procedure we assume signal and background migrate in identical fashion, because the background is dominated by peaking components. 

\begin{table}[htb]
\begin{center}
\caption{Migration matrix for $K^{0}_{\mathrm{S}}K^{+}K^{-}$ {\it vs}. $K^{+}K^{-}$ events when the $D\to K^{0}_{\rm S} K^{+}K^{-}$ Dalitz plot is divided into $\mathcal{N}=3$ bins.} \label{tab:migmatrix}
\setlength{\tabcolsep}{6.5pt}
\begin{tabular}{lcccccc} \hline\hline 
$i$ \hspace{0.5cm} & $U_{i,1}$ & $U_{i,2}$ & $U_{i,3}$ & $U_{i,-1}$ & $U_{i,-2}$ & $U_{i,-3}$ \\ \hline 
$\phantom{-}1$ & 0.968 & 0.020 & 0.001 & 0.011 &	0.000 & 0.000\\
\vspace{0.1cm}	
$\phantom{-}2$ & 0.036 & 0.967 & 0.001 &	0.000 &	0.001 & 0.003\\	
\vspace{0.1cm}
$\phantom{-}3$ & 0.007 & 0.001 & 0.992 &	0.000 &	0.000 & 0.000\\	
\vspace{0.1cm}
$-1$ & 0.010 & 0.000 & 0.000 & 0.972 & 0.018 &	0.000\\
\vspace{0.1cm}
$-2$ & 0.000 & 0.000 & 0.000 & 0.032 & 0.967 & 0.001 \\
\vspace{0.1cm}
$-3$ & 0.000 & 0.000 & 0.000 & 0.006 & 0.006 & 0.988 \\
\hline
\hline
\end{tabular} 
\end{center}
\end{table}  

The bin efficiency for each tag mode is evaluated from the signal MC sample. The signal MC yield in each bin is corrected for migration before calculating the efficiency. The bin efficiency is defined as the ratio of events reconstructed in each bin to the number of events generated. The bins are combined appropriately taking into account their symmetry (see Sec.~\ref{sec:formalism}) when estimating the efficiencies. The total DT efficiencies are given in Table~\ref{tab:yieldsandefficiencies}. In the case of $D\to K^{0}_{\rm S}K^{+}K^{-}$, the efficiencies vary between (12.43 $\pm$ 0.07)\% for $K^{0}_{\rm S}K^{+}K^{-}$ {\it vs.} $K^{-}\pi^{+}$ tags to (2.20 $\pm$ 0.03)\% for $K^{0}_{\rm S}K^{+}K^{-}$ {\it vs.} $K^{0}_{\rm S}\eta^{\prime}$ tags, whereas for $D\to K^{0}_{\rm L}K^{+}K^{-}$ the efficiency varies between ($15.85~\pm 0.08)$\% for $K^{0}_{\rm L}K^{+}K^{-}$ {\it vs.} $K^{-}\pi^{+}$ tags and (3.40 $\pm$ 0.04)\% for $K^{0}_{\rm S}K^{+}K^{-}$ {\it vs.} $K^{0}_{\rm L}K^{+}K^{-}$ tags. The uncertainty on the efficiency is related to the size of the MC sample. The bin efficiencies are used to calculate the expected yield for each tag mode as given in Eqs.~(\ref{eq:Mieqn}),~(\ref{eq:mijeqn}),~(\ref{eq:Miprimeeqn}) and~(\ref{eq:Mijprimeeqn}). 

Both pseudo-flavor DT yields with $F \in (K^{-}\pi^{+}$, $K^{-}\pi^{+}\pi^{0}$) have contamination from DCS decays whose contribution is enhanced compared to ST yields due to the quantum correlation between the $D^0\bar{D}^0$. Since these decays are used to determine $K^{(\prime)}_i$, the presence of this DCS contamination may bias the values~\cite{briere}. In order to correct for this effect, the yield in each bin is multiplied by a correction factor estimated using the decay model reported in Ref.~\cite{babarmodel}. The correction factors $f_{i}^{F}$ for $D^0\to K_{\rm S}^{0}K^{+}K^{-}$ against $F$ and $f^{F\prime}_{i}$ for $D^0\to K_{\rm L}^{0}K^{+}K^{-}$ against $F$ are given by
\begin{widetext}
\begin{eqnarray}
f^{F}_i&=&\frac{\int_i |f(m^2_+,m^2_-)|^2 dm^2_+dm^2_-}{\int_i (|f(m^2_+,m^2_-)|^2+(r_D^F)^2|f(m^2_-,m^2_+)|^2-2r^F_DR_F\mathcal{R}[e^{i\delta_D^F}f(m^2_+,m^2_-)f^*(m^2_-,m^2_+)])dm^2_+m^2_-}, \label{eq:dcscorrectionfactorforks} \\
f_i^{F\prime}&=&\frac{\int_i |f^\prime(m^2_+,m^2_-)|^2 dm^2_+dm^2_-}{\int_i (|f^\prime(m^2_+,m^2_-)|^2+(r_D^F)^2|f^\prime(m^2_-,m^2_+)|^2+2r^F_DR_F\mathcal{R}[e^{i\delta_D^F}f^\prime(m^2_+,m^2_-)f^{*\prime}(m^2_-,m^2_+)])dm^2_+dm^2_-},
\label{eq:dcscorrectionfactorforkl}
\end{eqnarray}
\end{widetext}
where $r_{D}^{F}$ is the ratio of the moduli of the DCS to CF amplitudes, for example $|\mathcal{A}(D^{0}\to K^{+}\pi^{-})/\mathcal{A}(D^{0}\to K^{-}\pi^{+})|$ for $K^{\pm}\pi^{\mp}$, and $\delta^{D}_{F}$ is the strong-phase difference between the DCS and CF amplitudes. The coherence factor, $R_{F}$ for flavor-mode $F$, accounts for the dilution in interference effects that arises when integrating over the phase space of multi-body decays \cite{coherencefactor}.  The values of the parameters used to determine the correction factors are listed in Table~\ref{tab:correctionfactorparameters}. The fraction of events in each bin $F_{i}^{(\prime)}$, defined in Eq. (\ref{eqn:relationkiandti}), is given in Table~\ref{tab:tivalues}. The $D^{0}\to K_{\rm L}^{0}K^{+}K^{-}$ amplitude model is developed by modifying the intermediate resonances of $D^{0}\to K_{\rm S}^{0}K^{+}K^{-}$ as presented in Ref.~\cite{chris_thesis}. Good agreement with the predicted values~\cite{jim} is observed for the results given in Table~\ref{tab:tivalues}. The uncertainties in the final result due to the correction factors are small and are treated as a systematic uncertainty. The DCS correction is not required for the $D^0\to K^{-}e^{+}\nu_{e}$ flavor-tag.

\begin{table}
\begin{center}
\caption{Values of the parameters used to calculate the DCS correction factors.}
\begin{tabular}{l c c c}
\hline
\hline
F & $r_{D}^{F}$ (\%) & $\delta_{D}^{F}$ ($^\circ$) & $R_{F}$\\
\hline
$K\pi$ & 5.86 $\pm$ 0.02 ~\cite{hfag} & $194.7^{+8.4}_{-17.6}$ ~\cite{hfag} & 1\\
$K\pi\pi^{0}$ & 4.47 $\pm$ 0.12 ~\cite{libbycoherencepaper} & $198^{+14}_{-15}$ ~\cite{libbycoherencepaper} & 0.81 $\pm$ 0.06 ~\cite{libbycoherencepaper}\\
\hline
\hline
\end{tabular}
\label{tab:correctionfactorparameters}
\end{center}
\end{table}

\begin{table}[ht!]
\begin{center}
\caption{Values of $F_{(-)i}$ and $F_{(-)i}^{\prime}$ (\%) measured from the flavor-tagged $D^{0}\to K^{0}_{\rm S}K^{+}K^{-}$ and $D^{0}\to K^{0}_{\rm L}K^{+}K^{-}$ data for the different number of bins $\mathcal{N}$.}
\setlength{\tabcolsep}{6.5pt}
\begin{tabular}{lcccc} 
\hline
\hline
$i$ & $F_i$ (\%) & $F_{-i}$ (\%) & $F_i^{\prime}$ (\%) & $F_{-i}^{\prime}$ (\%)\\
\multicolumn{5}{c}{$\mathcal{N}=2$ } \\ \hline
1 & $24.4\pm 1.7$  & $30.4\pm1.9$ & $23.5 \pm 1.2$ & $27.7 \pm 1.3$ \\
2 & $19.6\pm1.6$ & $25.6\pm1.9$ & $23.1 \pm 1.3$ & $25.6 \pm 1.3$\\
\multicolumn{5}{c}{$\mathcal{N}=3$ } \\ \hline
1 & $21.9\pm1.5$   & $27.7\pm1.8$ & $21.1 \pm 1.1$ & $25.1 \pm 1.2$ \\
2 & $21.3\pm1.7$  & $24.7\pm1.8$ & $22.6 \pm 1.3$ & $25.1 \pm 1.4$ \\
3 & \phantom{1}$1.3\pm0.4$ & \phantom{1}$3.1\pm0.5$ &\phantom{1}$2.8 \pm 0.3$ & \phantom{1}$3.3 \pm 0.4$\\
\multicolumn{5}{c}{$\mathcal{N}=4$} \\ \hline
1 & $21.1\pm1.5$ & $27.0\pm1.8$ & $19.5 \pm 1.0$ & $23.2 \pm 1.7$ \\
2 & \phantom{1}$6.5\pm0.9$ & \phantom{1}$3.6\pm0.6$ &\phantom{1}$7.2 \pm 0.7$ & \phantom{1}$4.1 \pm 0.5$\\
3 & $16.3\pm1.5$ & $22.4\pm1.8$ & $19.5 \pm 1.2$ & $23.0 \pm 1.3$\\
4 & \phantom{1}$0.5\pm0.2$ & \phantom{1}$2.6\pm0.5$ &\phantom{1}$0.9 \pm 0.2$ & \phantom{1}$2.6 \pm 0.3$\\ 
\hline
\hline
\end{tabular}
\label{tab:tivalues}
\end{center}
\end{table} 

\subsection{\boldmath Bin-by-bin background estimation}\label{sec:binbkgestimation}
In this section we explain the method used to estimate the peaking background. The amount of combinatorial background in each bin is estimated from the sideband-estimation methods described in Sec.~\ref{sec:dtyield}.

The peaking backgrounds are identified from the inclusive MC samples using the tool described in Ref.~\cite{topology}. The backgrounds to fully reconstructed tags are found to be negligible. However, all the $D\to K^{0}_{\rm L}X$ modes contain backgrounds from $D\to K^{0}_{\rm S}X$ modes, where the $\pi^{0}$ mesons from $K^{0}_{\rm S}\to \pi^{0}\pi^{0}$ decays are not reconstructed, so that the $K^{0}_{\rm S}$ is treated as a missing particle. The $D\to K^{0}_{\rm L}X$ and $D\to K^{0}_{\rm S}X$ decays are of opposite $CP$, hence the distribution of background events across the Dalitz plot is not the same as that for signal events. The level of these backgrounds varies between 2 to 4\% depending on the tag mode. The bin-by-bin background estimation using the inclusive MC sample is not reliable for two reasons. First, there can be a difference between the branching fraction in data and that assumed in the MC simulation. Second, the MC samples are not tuned to reflect the distributions of events across the Dalitz plot. Both these issues will result in an incorrect estimation of the bin-by-bin background. Hence, we use a combination of data and background MC samples to estimate the backgrounds.

Our method of peaking background estimation is as follows. We generate dedicated background MC samples corresponding to each type of background decay. The background retention efficiencies for these backgrounds are calculated for each bin. The expected yields are calculated using the values of $c_{i}$ and $s_{i}$ obtained from the previous measurement~\cite{jim} through the relations given in Eqs.~(\ref{eq:Mieqn}),~(\ref{eq:mijeqn}),~(\ref{eq:Miprimeeqn}) and~(\ref{eq:Mijprimeeqn}). The yields are multiplied by the retention efficiencies to obtain the expected background yields in data. Though most of the combinatorial background beneath the $D\to K^{0}_{\rm L}X$ signal decays are from non-resonant $D\to K^{+}K^{-}\pi^{0}\pi^{0}$ decays, these decays only contribute approximately 2\% of the background in the signal region. The contributions from continuum backgrounds and low-lying charmonium resonance decays are found to be negligible for all DT modes, which is expected given the tight kinematic criteria that can be imposed close to open-charm threshold. The expected background yields are not subtracted from signal yield but added to the expected signal yield in the fit as explained in Sec.~\ref{sec:cisiextraction}.

\begin{table*}[ht!]
\centering
\caption{Model-predicted values of $\Delta c_i$ and $\Delta s_i$ along with the uncertainties $\sigma_{\Delta c_i}$ and $\sigma_{\Delta s_i}$ for equal-$\Delta\delta_D$ binnings $\mathcal{N}=2,~3$ and $4$. The values are those reported in Ref.~\cite {chris_thesis}.}
\setlength{\tabcolsep}{6.5pt}
	\begin{tabular}{l c c c c c c c}
		\hline
		\hline
		$\mathcal{N}$ & $i$ & $c_{i}$ & $c_{i}^{\prime}$ & $\Delta c_i$ & $s_{i}$ & $s_{i}^{\prime}$ &  $\Delta s_i$ \\	
		\hline		 
		2 & 1 &  \phantom{0-}0.742 & \phantom{0-}0.768 & \phantom{0-}0.026 $\pm$ 0.017 & \phantom{0-}0.275 & 0.286 & \phantom{-} 0.011 $\pm$ 0.023\\
		  & 2 & $-$0.679 & $-$0.680 & $-$0.001 $\pm$ 0.036	& \phantom{0-}0.318 & 0.397 & \phantom{-} 0.079 $\pm$ 0.021\\ \cline{1-8}		      
		3 & 1 &  \phantom{0-}0.801 & \phantom{0-}0.827 & \phantom{0-}0.026 $\pm$ 0.014 & \phantom{0-}0.268 & 0.269 & \phantom{-} 0.001 $\pm$ 0.023\\
		  & 2 & $-$0.657 & $-$0.593 & \phantom{0-}0.064 $\pm$ 0.019 & \phantom{0-}0.411 & 0.435 & \phantom{-} 0.024 $\pm$ 0.010\\
		  & 3 & $-$0.043 & $-$0.680 & $-$0.637 $\pm$ 0.311 & $-$0.661 & 0.126 & \phantom{-} 0.787 $\pm$ 0.161\\
		  \cline{1-8}
		4 & 1 &  \phantom{0-}0.845 & \phantom{0-}0.864 & \phantom{0-}0.019 $\pm$ 0.011 & \phantom{0-}0.239 & 0.242 & \phantom{-} 0.003 $\pm$ 0.021\\
		  & 2 & $-$0.028 & \phantom{0-}0.095 & \phantom{0-}0.123 $\pm$ 0.029 & \phantom{0-}0.531 & 0.512 & $-$0.019 $\pm$ 0.022\\
		  & 3 & $-$0.804 & \phantom{0-}0.779 & \phantom{0-}0.025 $\pm$ 0.019 & \phantom{0-}0.332 & 0.382 & \phantom{0-}0.050 $\pm$ 0.015\\
		  & 4 &  \phantom{0-}0.232 & $-$0.718 & $-$0.950 $\pm$ 0.355 & \phantom{0-}0.738 & 0.262 & \phantom{0} 1.000 $\pm$ 0.254 \\  
		\hline
		\hline
	\end{tabular}
	\label{tab:deltacisiconstrainttable}	
\end{table*}

\begin{table*}[ht!]
	\centering
	\caption{Fit results for $c_{i}^{(\prime)}$ and $s_{i}^{(\prime)}$ for different $\mathcal{N}$. The first uncertainty is statistical and the second uncertainty is systematic.}
	\vspace{0.2cm}
	\setlength{\tabcolsep}{6.5pt}
	\begin{tabular}{c c c c c c c}
		\hline
		\hline
		$\mathcal{N}$ & $i$ & $c_{i}$ & $s_{i}$ & $c_{i}^{\prime}$ & $s_{i}^{\prime}$\\
		\hline
		2 & 1 & \phantom{0-}0.704 $\pm$ 0.034 $\pm$ 0.003 & $-$0.038 $\pm$ 0.144 $\pm$ 0.039 & \phantom{0-}0.730 $\pm$ 0.035 $\pm$ 0.003 & $-$0.028 $\pm$ 0.144 $\pm$ 0.039 \\
		  & 2 & $-$0.760 $\pm$ 0.040 $\pm$ 0.007 & \phantom{0-}0.590 $\pm$ 0.198 $\pm$ 0.085 & $-$0.785 $\pm$ 0.034 $\pm$ 0.006 & \phantom{0-}0.669 $\pm$ 0.198 $\pm$ 0.086 \\
		  \hline
		3 & 1 & \phantom{0-}0.724 $\pm$ 0.035 $\pm$ 0.003 & $-$0.037 $\pm$ 0.174 $\pm$ 0.049 & \phantom{0-}0.751 $\pm$ 0.036 $\pm$ 0.003 & $-$0.037 $\pm$ 0.174 $\pm$ 0.049\\
		  & 2 & $-$0.576 $\pm$ 0.050 $\pm$ 0.009 & \phantom{0-}0.616 $\pm$ 0.146 $\pm$ 0.047 & $-$0.512 $\pm$ 0.050 $\pm$ 0.009 & \phantom{0-}0.640 $\pm$ 0.146 $\pm$ 0.047 \\
		  & 3 & $-$0.174 $\pm$ 0.173 $\pm$ 0.040 & $-$0.669 $\pm$ 0.370 $\pm$ 0.119 & $-$0.382 $\pm$ 0.145 $\pm$ 0.040 & \phantom{0-}0.045 $\pm$ 0.384 $\pm$ 0.116\\
		 \hline
		4 & 1 & \phantom{0-}0.783 $\pm$ 0.034 $\pm$ 0.003 & $-$0.242 $\pm$ 0.173 $\pm$ 0.051 & \phantom{0-}0.802 $\pm$ 0.034 $\pm$ 0.003 & $-$0.239 $\pm$ 0.174 $\pm$ 0.051 \\
		  & 2 & $-$0.053 $\pm$ 0.106 $\pm$ 0.017 & \phantom{0-}0.306 $\pm$ 0.294 $\pm$ 0.125 & \phantom{0-}0.070 $\pm$ 0.106 $\pm$ 0.017 & \phantom{0-}0.286 $\pm$ 0.294 $\pm$ 0.124 \\
		  & 3 & $-$0.654 $\pm$ 0.057 $\pm$ 0.011 & \phantom{0-}0.659 $\pm$ 0.210 $\pm$ 0.059 & $-$0.630 $\pm$ 0.056 $\pm$ 0.011 & \phantom{0-}0.709 $\pm$ 0.210 $\pm$ 0.059 \\
		  & 4 & \phantom{0-}0.090 $\pm$ 0.208 $\pm$ 0.041 & $-$0.713 $\pm$ 0.387 $\pm$ 0.195 & $-$0.290 $\pm$ 0.201 $\pm$ 0.036 & \phantom{0-}0.122 $\pm$ 0.422 $\pm$ 0.206 \\   
		 \hline
		\hline
	\end{tabular}
	\label{tab:cisifittable}
\end{table*} 

\section{\boldmath Extraction of $c_{i}$ and $s_{i}$}\label{sec:cisiextraction}
{The uncorrected yields in related bins are combined according to the symmetry relations described in Sec.~\ref{sec:formalism}.} This procedure reduces the number of degrees of freedom in the fit. The quantities $K_{i}~(K_{i}^{\prime}$) and $c_{i},s_{i}$ ($c_{i}^{\prime},s_{i}^{\prime}$) for $D\to K^{0}_{\rm S}\pi^{+}\pi^{-}$ $(D\to K^{0}_{\rm L}\pi^{+}\pi^{-})$ are taken from Ref.~\cite{lei}. The values of $c_{i}^{(\prime)}$ and $s_{i}^{(\prime)}$ are obtained 
by minimizing the negative log likelihood expression 
\begin{eqnarray}
-2\ln\mathcal{L} & = &  -2\sum_i \ln P\left(N^\pm_i,\langle N^\pm_i\rangle\right)_{K^{0}_{\rm S}K^{+}K^{-},~CP} \nonumber \\
&& -2\sum_i \ln P\left(N^{\prime\pm}_i,\langle N^{\prime\pm}_i\rangle\right)_{K^{0}_{\rm L}K^{+}K^{-},~CP} \nonumber \\
&& -2\sum_{i,j} \ln P\left(N_{ij},\langle N_{ij}\rangle\right)_{K^{0}_{\rm S}K^{+}K^{-},~K^{0}_{\rm S}K^{+}K^{-}} \nonumber \\
&& -2\sum_{i,j} \ln P\left(N^{\prime}_{ij},\langle N^{\prime}_{ij}\rangle\right)_{K^{0}_{\rm S}K^{+}K^{-},~K^{0}_{\rm L}K^{+}K^{-}} \nonumber \\
&& -2\sum_{i,j} \ln P\left(N_{ij},\langle N_{ij}\rangle\right)_{K^{0}_{\rm S}K^{+}K^{-},K^{0}_{\rm S}\pi^{+}\pi^{-}} \nonumber \\
&& -2\sum_{i,j} \ln P\left(N^{\prime}_{ij},\langle N^{\prime}_{ij}\rangle\right)_{K^{0}_{\rm S}K^{+}K^{-},~K^{0}_{\rm L}\pi^{+}\pi^{-}} \nonumber \\
&& -2\sum_{i,j} \ln P\left(N^{\prime}_{ij},\langle N^{\prime}_{ij}\rangle\right)_{K^{0}_{\rm L}K^{+}K^{-},~K^{0}_{\rm S}\pi^{+}\pi^{-}} \nonumber \\
&& + \chi^2 \; .
\label{eq:kskkextract}
\end{eqnarray}
 Here, $\langle N \rangle$ is the expected migration-corrected yield in a particular bin whose measured yield is $N$. $P\left(N,\langle N \rangle\right)$ is the Poisson probability of observing a yield $N$ given the mean $\langle N \rangle$:
\begin{equation}
P\left(N,\langle N \rangle\right) = \frac{\langle N \rangle^{N}e^{-\langle N \rangle}}{N!} \; .
\end{equation}
If $\langle M \rangle$ represents the expected signal yield and $\langle B \rangle$ represents the expected background then $\langle N \rangle = \langle M \rangle + \langle B \rangle$. It is to be noted that in Eq.~(\ref{eq:kskkextract}) the comparison is between the uncorrected yield and the sum of expected signal and expected background in each bin. This method eliminates the possibility of unphysical negative bin yields arising from the subtraction of backgrounds from bins having low yields. 
The $\chi^{2}$ term in Eq.~(\ref{eq:kskkextract}) constrains the difference between the extracted values of $c_{i}~(s_{i})$ and $c'_{i}~(s'_{i})$ to lie within the uncertainties of the predicted differences $\Delta c_i~(\Delta s_i)$. The $\chi^2$ is defined as
\begin{equation}
\chi^2 = \sum_i \left(\frac{c_i'-c_i-\Delta c_i}{\sigma_{\Delta c_i}}\right)^2 + \sum_i \left(\frac{s_i'-s_i-\Delta 
s_i}{\sigma_{\Delta s_i}}\right)^2,
\end{equation} 
where $\Delta c_i = c'_{i,\textrm{BaBar}} - c_{i,\textrm{BaBar}} ~(\Delta s_i = s'_{i,\textrm{BaBar}} - s_{i,\textrm{BaBar}})$ is the predicted difference from the BaBar model~\cite{babarmodel} and $\sigma_{\Delta c_i}~(\sigma_{\Delta s_i})$ is the uncertainty on $\Delta c_i ~(\Delta s_i)$. The values of $\Delta c_i~(\Delta s_i)$ and $\sigma_{\Delta c_i} ~(\sigma_{\Delta s_i})$ are given in Table~\ref{tab:deltacisiconstrainttable}. 

Large values of $\Delta c_{i}$ and $\Delta s_{i}$ are observed in bins $i=3$ and $i=4$ in the $\mathcal{N}=3$ and $\mathcal{N}=4$ binnings, respectively. These bins correspond to the lobes on the Dalitz plot that encompass the neutral resonance $a_{0}(1450)^{0}$. The model has very small amplitudes in this region, hence a small difference between the $D\to K^{0}_{\rm S}K^{+}K^{-}$ and $D\to K^{0}_{\rm L}K^{+}K^{-}$ models is proportionally more significant. Consequently, the uncertainties are also large. Improvement in the precision due to the $\chi^{2}$ term is more significant for $s_{i}$ than $c_{i}$. The minimization of Eq.~(\ref{eq:kskkextract}) is performed using the {\sc MINUIT}~\cite{minuit} package. The value of parameters obtained from the fit are given in Table~\ref{tab:cisifittable}. 

The fitting procedure is validated using pseudodata samples generated by a standalone simulation. The bin yields and backgrounds are generated according to the relations given in Sec.~\ref{sec:formalism}. The effects of bin migration are also considered in the simulation. The bin yields are fluctuated assuming a Poisson distribution. The procedure is repeated 500 times to obtain a pull distributions for $c_i^{(\prime)}$ and $s_i^{(\prime)}$. The means and widths of the pulls in each bin are consistent with zero and one, respectively, indicating no bias and proper estimation of the uncertainties. 

\begin{figure*}[ht!] \centering
\includegraphics[width=7.7in,height=7.7in,keepaspectratio]{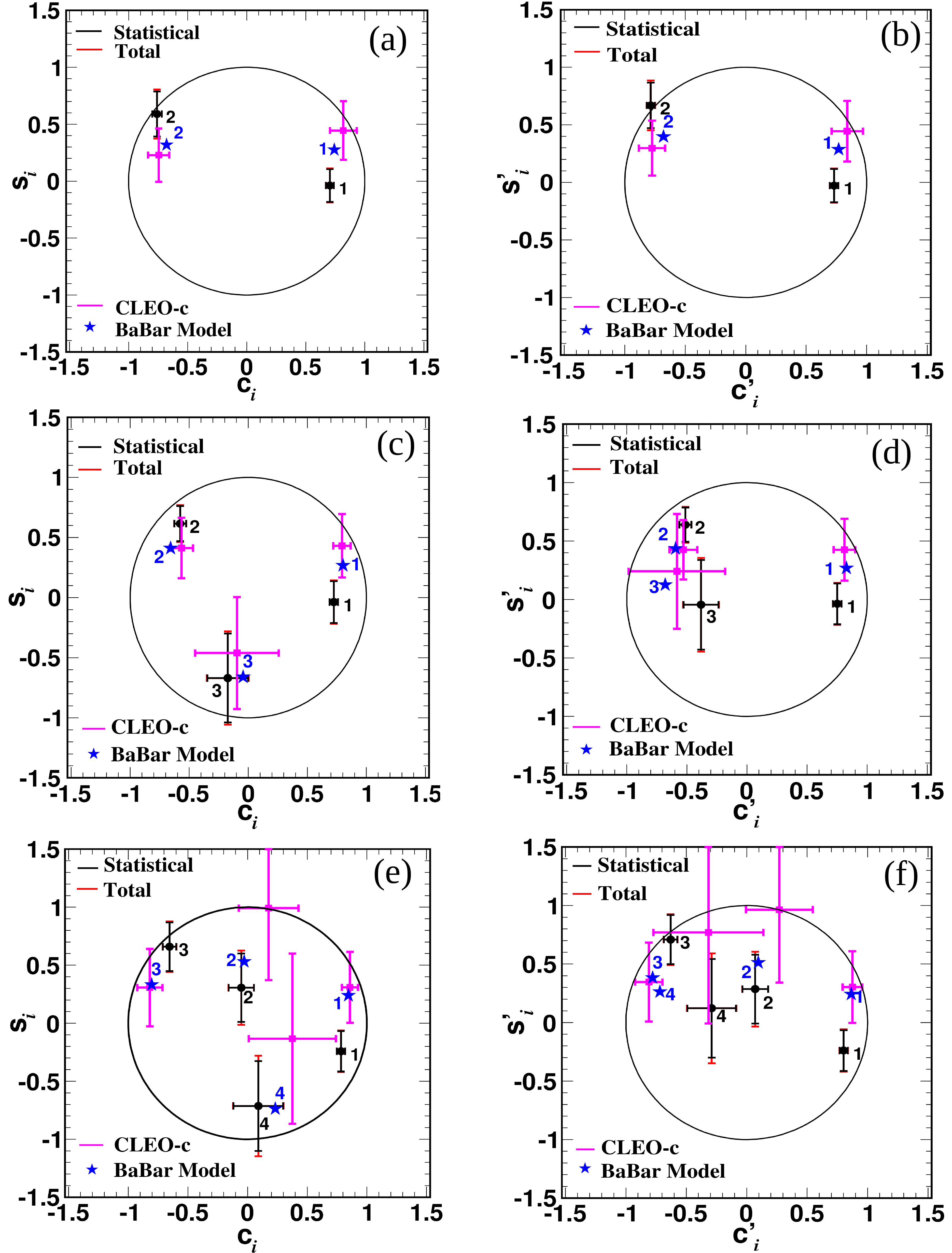}
\caption{Measured values of $c_{i}^{[\prime]}$ and $s_{i}^{[\prime]}$ for $\mathcal{N}=2$ (a) [(b)], $\mathcal{N}=3$ (c) [(d)], and $\mathcal{N}=4$ (e) [(f)] equal-$\Delta\delta_{D}$ bins are given by the black points with error bars. Also shown for comparison are the measurements reported by the CLEO Collaboration \cite{jim} (pink points with error bars) and the predictions of the model reported by the BaBar Collaboration \cite{babarmodel} (blue stars). The black circle corresponds to the allowed physical region $c_{i}^{2}+s_{i}^{2} = 1$.}
\label{fig:cisifits}
\end{figure*} 
\section{\boldmath Systematic uncertainties}\label{sec:systematics}

The evaluation of systematic uncertainties arising due to various inputs to the $c_{i}^{(\prime)}$ and $s_{i}^{(\prime)}$ fit are described in this section. In general, our method of evaluating systematic uncertainties uses a smearing procedure of the nominal input values within their uncertainties. This process is repeated 1000 times and the distribution of the resulting $c_{i}^{(\prime)}$ and $s_{i}^{(\prime)}$ values is fit with a Gaussian distribution. The width of the Gaussian distribution is then reported as the systematic uncertainty.
If there is a set of correlated input quantities assumed in our fit, we use a procedure based on the Cholesky decomposition of the covariance matrix to smear the value taking the correlation into account. The procedure involves generating a vector of correlated variables ${\bf X = \boldsymbol{\mu} + AZ}$, where $\boldsymbol{\mu}$ is a vector of the reported values of the input parameter, $\mathbf{Z}$ is a vector with  random values that follow a normal distribution, and {\bf A} is the decomposed matrix. The fit is repeated with the new value of ${\bf X}$. The systematic covariance matrix is calculated using the distributions of the $c_{i}^{(\prime)}$ and $s_{i}^{(\prime)}$ values as well as the correlations between them.

The systematic uncertainties related to the ST yields are evaluated as follows. 
First, a combined systematic uncertainty in the yields due to various assumptions made in the fit is estimated. For example, the endpoints of the ARGUS function are fixed in our fits. We rerun the fits by changing the endpoints by $\pm$0.5 MeV/$c^{2}$, which is the approximate uncertainty related to the beam-energy spread in the endpoint. The difference between the new value of the yield and the nominal value is taken as a systematic uncertainty in the ST yield. Other assumptions include the estimation of peaking backgrounds from MC simulations and the assumed branching fractions in partially reconstructed modes. The statistical uncertainties in the yields from the fits are added in quadrature with the systematic uncertainties to obtain the total uncertainty related to the ST fit.
The ST yield is smeared within the total uncertainty and the distribution of the resulting $c_{i}^{(\prime)}$ and $s_{i}^{(\prime)}$ values is obtained as laid out above. The systematic uncertainties due to the ST yield are found to be small. The systematic uncertainty due to the factors used to correct for the effect of charm mixing is found to be negligible. 
Furthermore, uncertainties related to the absolute efficiency do not affect the results due to the use of yield ratios and normalization constants fit to data.

The systematic uncertainty due to the finite statistics of the flavor-tag yields $K_{i}$ are evaluated by  smearing the input values assuming a Gaussian distribution around the nominal value with width equal to the uncertainty. The systematic uncertainty due to the flavor-tag yield of the $D\to K^{0}_{\rm S}\pi^{+}\pi^{-}$ decay is estimated using the covariance matrix reported in the Ref.~\cite{lei}. The uncertainties due to flavor-tag yields are small due to the large yields compared to those for $CP$ and $D\to K^0h^+h^-$ tags. 

In our fit the values of $c_{i}^{(\prime)}$ and $s_{i}^{(\prime)}$ of $D\to K^{0}_{\rm S, L}\pi^{+}\pi^{-}$ are fixed to the values reported in Ref.~\cite{lei}. The covariance matrix used to smear the values is also taken from Ref.~\cite{lei}. This uncertainty is the dominant systematic uncertainty.

The value of the total number of
$D^0\bar{D}^0$ pairs is fixed in our fits. This information is used to normalize the bin yields of $D\to K^{0}_{\rm S, L}K^{+}K^{-}$ tagged by $D\to K^{0}_{\rm S, L}h^{+}h^{-}$ decays. The related systematic uncertainty is evaluated by smearing the value of $N_{D^{0}\bar{D}^{0}}$ within its uncertainty assuming a Gaussian distribution. Since the value of $N_{D^0\bar{D}^0}$ pairs is measured precisely~\cite{besNDDbar}, the systematic uncertainty due to this input is negligible.  

Signal MC samples are widely used in this analysis for various purposes such as constructing migration matrices and determining the selection efficiencies. The systematic uncertainties due to the finite size of the MC samples are evaluated by smearing the matrix elements within their uncertainties assuming a Gaussian distribution taking correlations into account. Any systematic uncertainty due to the incorrect MC modelling is cancelled since we use ratios of ST and DT yields in computing the values of $\langle M\rangle$.

The effect of bin-by-bin peaking background from $D\to K^{0}_{\rm S} h^{+}h^{-}$ decays in the $D\to K^{0}_{\rm L} h^{+}h^{-}$ signal sample is estimated using the $c_{i}$ and $s_{i}$ values reported in Ref.~\cite{jim}. To estimate the systematic uncertainty from the background parametrization, the estimated amount of background is varied by a Gaussian function within its statistical uncertainty. The level of the background with respect to the partially reconstructed tags is larger than that with respect to the fully reconstructed tags, hence the systematic uncertainties due to the background parametrization for partially reconstructed tags are larger than those for the fully reconstructed tags. Since the backgrounds are only identified but not estimated from MC simulations, there is no systematic uncertainty arising due to any difference in branching fractions between data and MC simulations.

The systematic uncertainties due to the assumed values of the DCS correction factors to the flavor-tag yields $f_{i}^{F(\prime)}$ are estimated by smearing the correction factors within their uncertainties assuming a Gaussian distribution. The $f_{i}^{F(\prime)}$ uncertainties are small, hence the associated systematic uncertainties on the values of $c_i^{(\prime)}$ and $s_i^{(\prime)}$ are also small.

An example of the individual contributions to the systematic uncertainties for the $\mathcal{N}=3$ equal-$\Delta\delta_D$ binning is shown in Table~\ref{tab:systsummary3bin}; corresponding tables for the $\mathcal{N}=2$ and $4$ binning schemes are given in App.~\ref{app:systematics}. The systematic uncertainties are significantly smaller than statistical uncertainties for all binning schemes. Appendix~\ref{app:correlations} contains the statistical and systematic correlation matrices for the $\mathcal{N}=2,~3$ and $4$ binning schemes. The final results for $c_{i},s_{i},c_{i}^{\prime}$ and $s_{i}^{\prime}$ are shown in Fig.~\ref{fig:cisifits}.

\begin{table*}[ht!]
	\centering
	\caption{Summary of the contributions to the systematic uncertainty for the  $\mathcal{N}=3$ equal-$\Delta\delta_D$ binning.}	
	\setlength{\tabcolsep}{6.5pt}
	\begin{tabular}{lcccccccccccc}
	\hline
	\hline
	Systematic & $c_{1}$ & $c_{2}$ & $c_{3}$ & $s_{1}$ & $s_{2}$ & $s_{3}$ & $c'_{1}$ & $c'_{2}$ & $c'_{3}$ & $s'_{1}$ & $s'_{2}$ & $s'_{3}$\\
	\hline
	ST yield & 0.001 & 0.005 & 0.007 & 0.001 & 0.002 & 0.001 & 0.001 & 0.005 & 0.002 & 0.001 & 0.002 & 0.001\\
	$K_{i}^{(\prime)}$ statistics & 0.001 & 0.006 & 0.033 & 0.006 & 0.001 & 0.010 & 0.001 & 0.006 &  0.033 & 0.006 & 0.001 & 0.010\\
	$K^{0}\pi^{+}\pi^{-}(c_{i}^{(\prime)}$, $s_{i}^{(\prime)})$ & 0.002 & 0.003 & 0.010 & 0.045 & 0.044 & 0.085 & 0.002 & 0.003 & 0.004 & 0.045 & 0.044 & 0.083\\
	$K^{0}\pi^{+}\pi^{-}(K_{i}^{(\prime)})$ & 0.001 & 0.001 & 0.003 & 0.009 & 0.015 & 0.018 & 0.001 & 0.001 & 0.001 & 0.009 & 0.015 & 0.018\\
	$N_{D\bar{D}}$ & 0.000 & 0.000 & 0.000 & 0.001 & 0.000 & 0.000 & 0.000 & 0.000 & 0.000 & 0.001 & 0.000 & 0.001\\	
	MC statistics & 0.001 & 0.003 & 0.013 & 0.000 & 0.001 & 0.000 & 0.001 & 0.003 & 0.020 & 0.000 & 0.001 & 0.000\\
	Background & 0.001 & 0.002 & 0.012 & 0.015 & 0.009 & 0.082 & 0.001 & 0.002 & 0.008 & 0.015 & 0.009 & 0.079\\
	DCS correction &  0.000 & 0.000 & 0.000 & 0.000 & 0.000 & 0.000 & 0.000 & 0.000 & 0.000 & 0.000 & 0.000 & 0.000\\
	\hline
	Statistical & 0.035 & 0.050 & 0.173 & 0.174 & 0.146 & 0.370 & 0.036 & 0.050 & 0.145 & 0.174 &  0.146 & 0.384\\
	Total systematic & 0.003 & 0.009 & 0.040 & 0.049 & 0.047 & 0.119 & 0.003 & 0.009 & 0.040 & 0.049 & 0.047 & 0.116 \\
	\hline
	Total & 0.035 & 0.058 & 0.178 & 0.181 & 0.154 & 0.389 & 0.036 & 0.051 & 0.150 & 0.181 & 0.154 & 0.401 \\
	\hline
	\hline
\end{tabular}
\label{tab:systsummary3bin}
\end{table*} 

\section{\boldmath Impact of $c_{i}, s_{i}$ on measurement of $\gamma$}\label{sec:impact}
The values of $c_{i}$ and $s_{i}$ are used as an input to the model-independent determination of $\gamma$ using the $B^{-}\to DK^{-},~D\to~K^{0}_{\rm S}K^{+}K^{-}$ decay. The uncertainties on the measured values of $c_{i}$ and $s_{i}$ introduce a related systematic uncertainty on the measured value of $\gamma$, which we here estimate through a  pseudodata simulation study. 

\begin{figure*}[ht!] \centering
\includegraphics[width=7.0in,height=7.0in,keepaspectratio]{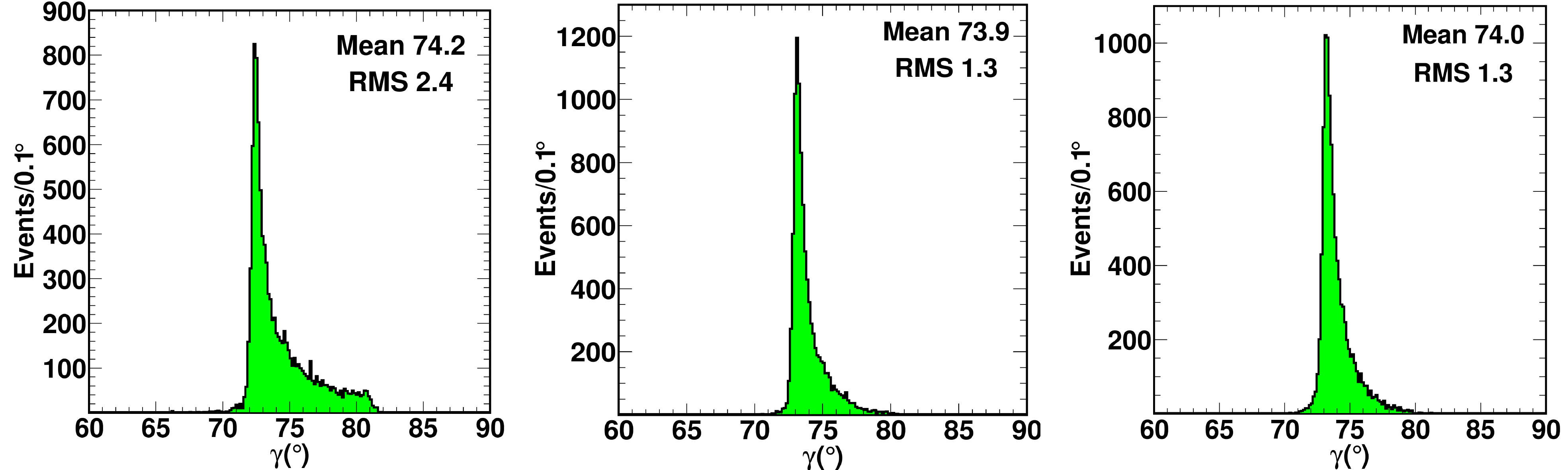}
\caption{Distribution of $\gamma$ obtained from pseudodata experiments for $\mathcal{N}=2$ (left), $\mathcal{N}=3$ (middle) and $\mathcal{N}=4$ (right) binning schemes.}
\label{fig:gammadistributiontoymc}
\end{figure*}

We simulate the decay rate of $B^{-}\to DK^{-},~D\to~K^{0}_{\rm S}K^{+}K^{-}$ within the Dalitz plot bins using the relation in Eq.~(\ref{eq:bdecay_rate}). In the simulation, the input values of $c_{i}, s_{i}, T_{i}$ and $T_{-i}$ are set to those reported here. The values of $r_{B}$, $\delta_{B}$ and $\gamma$ are taken to be 0.103, $136.9^{\circ}$ and $73.5^{\circ}$,  respectively (see Ref.~\cite{hfag}). In order to reduce the uncertainty due to the statistics of the $B$ decay sample to a negligible level, the pseudodata sample size is set to 5$\times 10^{6}$ events. The bin yields are fluctuated according to a Poisson distribution to produce a new bin yield, $N^{I}$. The expected bin yield, $N^{E}$, is calculated using $c_{i}$ and $s_{i}$ values obtained by smearing the measured values by their uncertainties, assuming a Gaussian distribution and taking the correlations into account. The best fit values of $r_{B}, \delta_{B}$ and $\gamma$ are extracted by minimizing $\chi^{2} = \sum_{i}(N^{I}_{i} - N^{E}_{i})^{2}/N^{E}_{i}$. The pseudodata simulations are repeated 10000 times and the resulting $\gamma$ distributions for $\mathcal{N}=2,~3$ and 4 bins are shown in Fig.~\ref{fig:gammadistributiontoymc}.

The distributions of all parameters ($r_{B}, \delta_{B}$ and $\gamma$) are found to be  asymmetric. Using the root mean square (RMS) of the distribution, we estimate the uncertainties on $\gamma$  to be 2.3$^{\circ}$, 1.3$^{\circ}$ and 1.3$^{\circ}$ for the schemes with $\mathcal{N}=2,~3$ and 4 bins, respectively. We also estimate an asymmetric uncertainty by integrating 16\% of the distribution in the lower and upper tails to work out a 68\% confidence level; the asymmetric uncertainties on $\gamma$ are 
%
$^{+2.4^{\circ}}_{-1.8^{\circ}}$, $^{+1.0^{\circ}}_{-0.9^{\circ}}$ and $^{+0.9^{\circ}}_{-0.9^{\circ}}$
for schemes with $\mathcal{N}=2, 3$ and 4 bins, respectively. Better sensitivity for $\mathcal{N}=3$ and $4$ compared to $\mathcal{N}=2$ is observed. This is due to the fact that dividing the data into more bins is a better approximation of the unbinned case. The lack of improvement in sensitivity while going from $\mathcal{N}=3$ to $\mathcal{N}=4$ is due to the nature of the $D^0 \to K^{0}_{\rm S}K^{+}K^{-}$ Dalitz plot. The dominant resonances contributing to $D^0\to K^{0}_{\rm S}K^{+}K^{-}$ are $D^0\to K^{0}_{\rm S}\phi$ and $D^0\to K^{0}_{\rm S}a(980)^{0}$,
which are both located close to the $M^{2}_{K^{+}K^{-}}$ kinematic limit. In a binning based on equal-$\Delta\delta_{D}$ regions these bins are always enclosed by a similar pair of bins and the new pair of bins always encloses a region with a low density of events, as can be seen in Fig.~\ref{fig:kskkbins_model}. 

Biases of up to $0.7^\circ$ are observed in the central values of $\gamma$ for all binning schemes. A  bias was reported in the previous analysis as well~\cite{jim}. The bias is smaller with our measurement, but non-negligible given the size of the pseudodata sample used. In order to  investigate the source of the bias, we rerun the pseudodata experiments ignoring any pairs of $c_{i}$ and $s_{i}$ values that lie outside the physical region  given by $c_{i}^{2}+s_{i}^{2} < 1$; the bias is still observed, which is due to the non-Gaussian nature of the truncated $c_{i}$ and $s_{i}$ distributions. Therefore, instead of removing the unphysical values the uncertainties on the $c_{i}$ and $s_{i}$ are artificially reduced by a factor of three; in this case no observable bias is seen in any of the extracted parameters. Hence we conclude that the bias stems from some pairs of $c_{i}$  and $s_{i}$ values lying outside the physical region. Therefore, future measurements with a larger $\psi(3770)$ sample \cite{bes3whitepaper} are likely to reduce or remove this bias.

\section{\boldmath Summary}\label{sec:summary}
Using a sample of $\psi(3770)$ data corresponding to an integrated luminosity of 2.93  fb$^{-1}$ collected by the BESIII detector, we report a measurement of the strong-phase difference parameters for $D\to K^{0}_{\rm S,L}K^{+}K^{-}$ decays with the best precision to date. The results presented here are an important input to model-independent measurements of the CKM angle $\gamma$ using the BPGGSZ method. These values are also essential for the model-independent determination of charm-mixing parameters and in the search for $CP$ violation in $D^{0}\to K^{0}_{\rm S}K^{+}K^{-}$ decays \cite{lhcb}. We note that the statistical uncertainties limit the precision of our measurements. Therefore, it is desirable to collect larger  $\psi(3770)$ data sets in future \cite{bes3whitepaper}.

The major source of systematic uncertainty is due to the input strong-phase difference parameters of $D\to K^{0}_{\rm S,L}\pi^{+}\pi^{-}$ decays \cite{lei}. Significant systematic uncertainties also arise from the background parametrization of the partially reconstructed $D\to K^{0}_{\rm L}X$ decay modes. Both of  these uncertainties depend on the size of the charm sample available. 

Good agreement with the previous measurements by the CLEO Collaboration~\cite{jim} is obtained in all bins.  Hence, we have performed a combination of the BESIII and CLEO results, which is reported in App.~\ref{app:combination}. The predictions from the BaBar model~\cite{babarmodel} lie within one to two standard deviations from our values. We have recently reported an amplitude model and branching fraction for $D^0\to K_{\rm S}^{0}K^+K^-$ decays \cite{bib:bes3k0skk}; the measured values of $c_i$ and $s_i$ are also in agreement with this model. The estimated uncertainties on $\gamma$ arising from the uncertainties on the measured values of $c_i$ and $s_i$ are 2.3$^\circ$, 1.3$^\circ$ and 1.3$^\circ$ for $\mathcal{N}=2$, $\mathcal{N}=3$ and $\mathcal{N}=4$ equal-$\Delta\delta_{D}$ binning, respectively.

\section{\boldmath Acknowledgments}\label{acknowledgements}
The BESIII Collaboration thanks the staff of BEPCII and the IHEP computing center for their strong support. This work is supported in part by National Key Basic Research Program of China under Contract No. 2015CB856700; National Natural Science Foundation of China (NSFC) under Contracts Nos. 11625523, 11635010, 11735014, 11822506, 11835012, 11935015, 11935016, 11935018, 11961141012; the Chinese Academy of Sciences (CAS) Large-Scale Scientific Facility Program; Joint Large-Scale Scientific Facility Funds of the NSFC and CAS under Contracts Nos. U1732263, U1832207; CAS Key Research Program of Frontier Sciences under Contracts Nos. QYZDJ-SSW-SLH003, QYZDJ-SSW-SLH040; 100 Talents Program of CAS; INPAC and Shanghai Key Laboratory for Particle Physics and Cosmology; ERC under Contract No. 758462; German Research Foundation DFG under Contracts Nos. 443159800, Collaborative Research Center CRC 1044, FOR 2359, FOR 2359, GRK 214; Istituto Nazionale di Fisica Nucleare, Italy; Ministry of Development of Turkey under Contract No. DPT2006K-120470; National Science and Technology fund; Olle Engkvist Foundation under Contract No. 200-0605; STFC (United Kingdom); The Knut and Alice Wallenberg Foundation (Sweden) under Contract No. 2016.0157; The Royal Society, UK under Contracts Nos. DH140054, DH160214; The Swedish Research Council; U. S. Department of Energy under Contracts Nos. DE-FG02-05ER41374, DE-SC-0012069.

\appendix
\section{Additional tables of systematic uncertainties}
\label{app:systematics}
The contributions to the systematic uncertainty of the $c_{i}^{(\prime)}$ and $s_{i}^{(\prime)}$ measurements for the $\mathcal{N}=2$ and $\mathcal{N}=4$ binning schemes are given in Tables~\ref{tab:systsummary2bin} and \ref{tab:systsummary4bin}, respectively. These tables complement Table~\ref{tab:systsummary3bin} for the $\mathcal{N}=3$ binning scheme, which is given in Sec.~\ref{sec:systematics}.
\begin{table*}[ht!]
	\centering
	\caption{Summary of the contributions to the systematic uncertainty for the  $\mathcal{N}=2$ equal-$\Delta\delta_D$ binning.}
	\setlength{\tabcolsep}{6.5pt}
	\begin{tabular}{l c c c c c c c c}
	\hline
	\hline
	Systematic & $c_{1}$ & $c_{2}$ & $s_{1}$ & $s_{2}$ & $c'_{1}$ & $c'_{2}$ & $s'_{1}$ & $s'_{2}$ \\
	\hline
	ST yield & 0.002 & 0.003 & 0.000 & 0.000 & 0.002 & 0.001 & 0.000 & 0.000\\
	$K_{i}^{(')}$ statistics & 0.000 & 0.003 & 0.004 & 0.005 & 0.001 & 0.002 & 0.004 & 0.004\\
	$K^{0}\pi^{+}\pi^{-} (c_{i}^{(\prime)}$, $s_{i}^{(\prime)})$ & 0.001 & 0.002 & 0.037 & 0.075 & 0.001 & 0.002 & 0.037 & 0.076 \\
	$K^{0}\pi^{+}\pi^{-}(K_{i}^{(\prime)})$ & 0.000 & 0.001 & 0.007 & 0.033 & 0.000 & 0.000 & 0.006 & 0.033\\	
	$N_{D\bar{D}}$ & 0.000 & 0.000 & 0.001 & 0.001 & 0.000 & 0.000 & 0.001 & 0.001\\
	MC statistics & 0.001 & 0.002 & 0.000 & 0.000 & 0.001 & 0.003 & 0.000 & 0.000\\
	Background & 0.002 & 0.003 & 0.011 & 0.022 & 0.002 & 0.003 & 0.011 & 0.022\\
	DCS correction & 0.000 & 0.000 & 0.000 & 0.001 & 0.000 & 0.000 & 0.000 & 0.001\\
	\hline
	Stat & 0.034 & 0.040 & 0.144 & 0.198 & 0.035 & 0.034 & 0.144 & 0.198\\
	Syst total & 0.003 & 0.007 & 0.039 & 0.085 & 0.003 & 0.006 & 0.039 & 0.086\\
	\hline
	Total & 0.034 & 0.041 & 0.149 & 0.215 & 0.035 & 0.035 & 0.149 & 0.216\\
	\hline
	\hline
\end{tabular}
\label{tab:systsummary2bin}
\end{table*} 

\begin{table*}[ht!]\scriptsize
	\centering
	\caption{Summary of the contributions to the systematic uncertainty for the  $\mathcal{N}=4$ equal-$\Delta\delta_D$ binning.}
	\setlength{\tabcolsep}{3.5pt}
	\begin{tabular}{l c c c c c c c c c c c c c c c c}
	\hline
	\hline
	Systematic & $c_{1}$ & $c_{2}$ & $c_{3}$ & $c_{4}$ & $s_{1}$ & $s_{2}$ & $s_{3}$ & $s_{4}$ & $c'_{1}$ & $c'_{2}$ & $c'_{3}$ & $c'_{4}$ & $s'_{1}$ & $s'_{2}$ & $s'_{3}$ & $s'_{4}$\\
	\hline
	ST yield & 0.001 & 0.003 & 0.006 & 0.004 & 0.002 & 0.006 & 0.002 & 0.003 & 0.001 & 0.003 & 0.006 & 0.001 & 0.002 & 0.006 & 0.002 & 0.000\\
	$K_{i}^{(\prime)}$ statistics & 0.002 & 0.008 & 0.007 & 0.010 & 0.002 & 0.009 & 0.005 & 0.026 & 0.002 & 0.008 & 0.007 & 0.010 & 0.002 & 0.009 & 0.005 & 0.026\\
	$K^{0}\pi^{+}\pi^{-}(c_{i}, s_{i})$ & 0.002 & 0.003 & 0.003 & 0.015 & 0.048 & 0.089 & 0.053 & 0.091 & 0.002 & 0.003 & 0.003 & 0.008 & 0.048 & 0.089 & 0.053 & 0.087\\
	$K^{0}\pi^{+}\pi^{-}(K_{i}^{(\prime)})$ & 0.001 & 0.001 & 0.001 & 0.003 & 0.009 & 0.013 & 0.017 & 0.016 & 0.001 & 0.001 & 0.001 & 0.001 & 0.009 & 0.013 & 0.017 & 0.016\\
	$N_{D\bar{D}}$ & 0.001 & 0.000 & 0.001 & 0.001 & 0.001 & 0.003 & 0.004 & 0.000 & 0.001 & 0.000 & 0.001 & 0.000 & 0.001 & 0.003 & 0.004 & 0.000\\
	MC statistics & 0.000 & 0.013 & 0.003 & 0.015 & 0.002 & 0.006 & 0.002 & 0.001 & 0.000 & 0.012 & 0.003 & 0.026 & 0.001 & 0.006 & 0.002 & 0.001\\
	Background & 0.001 & 0.007 & 0.003 & 0.033 & 0.013 & 0.086 & 0.019 & 0.170 & 0.001 & 0.007 & 0.003 & 0.018 & 0.013 & 0.085 & 0.019 & 0.182 \\
	DCS correction & 0.000 & 0.000 & 0.000 & 0.001 & 0.001 & 0.000 & 0.001 & 0.000 &  0.000 & 0.000 & 0.000 & 0.002 & 0.000 & 0.000 & 0.001 & 0.000\\
	\hline
	Stat & 0.034 & 0.106 & 0.057 & 0.208 & 0.173 & 0.294 & 0.210 & 0.387 & 0.034 & 0.106 & 0.056 & 0.201 & 0.174 & 0.294 & 0.210 & 0.422\\
	Syst total & 0.003 & 0.017 & 0.011 & 0.041 & 0.051 & 0.125 & 0.059 & 0.195 & 0.003& 0.017 & 0.011 & 0.036 & 0.051 & 0.124 & 0.059 & 0.206\\
	\hline
	Total & 0.034 & 0.107 & 0.058 & 0.212 & 0.180 & 0.320 & 0.218 & 0.433 & 0.034 & 0.107 & 0.057 & 0.204 & 0.181 & 0.319 & 0.218 & 0.469\\
	\hline
	\hline
\end{tabular}
\label{tab:systsummary4bin}
\end{table*} 

\section{Statistical and systematic covariance matrices}
\label{app:correlations}
We report the statistical (systematic) covariance matrices related to the measurements for the $\mathcal{N}=2,~3$ and 4 binning schemes in Tables~\ref{tab:statcorrmat2bin} (\ref{tab:systcorrmat2bin}), \ref{tab:statcorrmat3bin} (\ref{tab:systcorrmat3bin}) and \ref{tab:statcorrmat4bin} (~\ref{tab:systcorrmat4bin}), respectively.

\begin{table*}[ht!]
	\centering
	\caption{Statistical correlation matrix (\%) for the $K^{0}_{\mathrm{S}}K^{+}K^{-}$ equal-$\Delta\delta_{D}$ $\mathcal{N}=2$ binning.}	
	\label{tab:statcorrmat2bin}
	\setlength{\tabcolsep}{2.5pt}
	\begin{tabular}{l d d d d d d d d }
	\hline
	\hline
	 & \multicolumn{1}{c}{$c_{2}$} & \multicolumn{1}{c}{$s_{1}$} & \multicolumn{1}{c}{$s_{2}$} & \multicolumn{1}{c}{$c_{1}^{\prime}$} & \multicolumn{1}{c}{$c_{2}^{\prime}$} & \multicolumn{1}{c}{$s_{1}^{\prime}$} & \multicolumn{1}{c}{$s_{2}^{\prime}$} \\
	 \hline
	 $c_{1}$ &  4.8 & 2.9 & -0.2 & 94.1 & 3.0 & 2.9 &-0.2 \\
	 $c_{2}$ & & -1.2 & -1.5 & 5.1 & 63.2 & -1.2 & -1.5 \\
	 $s_{1}$ & & & -0.4 & 2.8 & -0.8 & 99.4 & 0.4  \\
	 $s_{2}$ & & & & -0.2 & -1.6 & -0.3 & 99.5 \\
	 $c_{1}^{\prime}$ & & & & &  3.2 & 2.8 & -0.2 \\
	 $c_{2}^{\prime}$ & & & & & & -0.8 & 1.6\\
	 $s_{1}^{\prime}$& & & & & & &  0.3\\
	 \hline	
	 \hline	 
\end{tabular}
\end{table*} 
\begin{table*}[ht!]
	\centering
	\caption{Systematic correlation (\%) matrix for $K^{0}_{\mathrm{S}}K^{+}K^{-}$ equal-$\Delta\delta_{D}$ $\mathcal{N}=2$ binning.}	
	\label{tab:systcorrmat2bin}
	\setlength{\tabcolsep}{2.5pt}
	\begin{tabular}{l d d d d d d d d }
	\hline
	\hline
	 & \multicolumn{1}{c}{$c_{2}$} & \multicolumn{1}{c}{$s_{1}$} & \multicolumn{1}{c}{$s_{2}$} & \multicolumn{1}{c}{$c_{1}^{\prime}$} & \multicolumn{1}{c}{$c_{2}^{\prime}$} & \multicolumn{1}{c}{$s_{1}^{\prime}$} & \multicolumn{1}{c}{$s_{2}^{\prime}$} \\
	 \hline
	 $c_{1}$ &  36.6& -1.3 & -0.1 & 90.0& 27.3& -1.3 & -0.1 \\
	 $c_{2}$ & & -5.3 & -8.4 & 26.7& 58.4& -5.4 & -8.4	\\
	 $s_{1}$ & & & -21.1 &  1.6& 12.7& 99.6 & -21.2 \\
	 $s_{2}$ & & & & 0.7& -10.9 & -21.0 & 90.0\\
	 $c_{1}^{\prime}$ & & & & &  33.4&  -1.6 &  -0.7\\
	 $c_{2}^{\prime}$ & & & & & & 12.6& -11.1	\\
	 $s_{1}^{\prime}$& & & & & & &  -21.1 \\
	 \hline	
	 \hline	 
\end{tabular}
\end{table*}	

\begin{table*}[ht!]
	\centering
	\caption{Statistical correlation matrix (\%) for $K^{0}_{\mathrm{S}}K^{+}K^{-}$ equal-$\Delta\delta_{D}$ $\mathcal{N}=3$ binning.}
	\label{tab:statcorrmat3bin}	
	\setlength{\tabcolsep}{2.5pt}
		\begin{tabular}{l d d d d d d d d d d d d }
	\hline
	\hline
	    & \multicolumn{1}{c}{$c_{2}$} &
	    \multicolumn{1}{c}{$c_{3}$} &
        \multicolumn{1}{c}{$s_{1}$} & 
        \multicolumn{1}{c}{$s_{2}$} & 
        \multicolumn{1}{c}{$s_{3}$} &
        \multicolumn{1}{c}{$c_{1}^{\prime}$} & \multicolumn{1}{c}{$c_{2}^{\prime}$} &
    \multicolumn{1}{c}{$c_{3}^{\prime}$} &
    \multicolumn{1}{c}{$s_{1}^{\prime}$} & \multicolumn{1}{c}{$s_{2}^{\prime}$} &
    \multicolumn{1}{c}{$s_{3}^{\prime}$} \\
	 \hline
	 $c_{1}$ & 1.3&  1.8& 1.2& 0.1& 0.0& 9.7& 1.2& 0.4& -1.2& 0.1& 0.0\\
	 $c_{2}$ &    &  0.7& 0.1& 9.9& 1.5& 1.3& 96.5& 0.1& 0.1&  9.9& 1.3\\
	 $c_{3}$ &    &    & -1.4& -1.3& 0.9& 1.8& 0.7& 24.3& -1.4& -1.3& 0.8\\
	 $s_{1}$ &    &      &  &-5.5& -7.2& 1.1& 0.1& -0.3& 99.7& -5.5& -6.6\\
	 $s_{2}$ &    &      &    &  & 11.7& 0.1& 9.5& -0.4& -5.5&  99.9& 10.7\\
	 $s_{3}$ &    &      &    &    &  & 0.0& 1.4& -0.3& -7.2& 11.7& 91.3 \\
	 $c_{1}^{\prime}$ &    &      &    &    &    &  &  1.3& 0.4& -1.1& 0.1& 0.0 \\
	 $c_{2}^{\prime}$ &    &      &    &    &    &    &  & 0.2& 0.1& 9.4& 1.3 \\
	 $c_{3}^{\prime}$ &    &      &    &    &    &    &    &  &-0.3& -0.4& -0.4\\
	 $s_{1}^{\prime}$ &    &      &    &    &    &    &    &    &  & -5.5& -6.6 \\
	 $s_{2}^{\prime}$ &    &      &    &    &    &    &    &    &    &  & 10.7\\
     \hline
	 \hline	 
\end{tabular}
\end{table*} 
\begin{table*}[ht!]
	\centering
	\caption{Systematic correlation matrix (\%) for $K^{0}_{\mathrm{S}}K^{+}K^{-}$ equal-$\Delta\delta_{D}$ $\mathcal{N}=3$ binning.}
	\label{tab:systcorrmat3bin}	
	\setlength{\tabcolsep}{2.5pt}
		\begin{tabular}{l d d d d d d d d d d d d }
	\hline
	\hline
	    & \multicolumn{1}{c}{$c_{2}$} &
	    \multicolumn{1}{c}{$c_{3}$} &
        \multicolumn{1}{c}{$s_{1}$} & 
        \multicolumn{1}{c}{$s_{2}$} & 
        \multicolumn{1}{c}{$s_{3}$} &
        \multicolumn{1}{c}{$c_{1}^{\prime}$} & \multicolumn{1}{c}{$c_{2}^{\prime}$} &
    \multicolumn{1}{c}{$c_{3}^{\prime}$} &
    \multicolumn{1}{c}{$s_{1}^{\prime}$} & \multicolumn{1}{c}{$s_{2}^{\prime}$} &
    \multicolumn{1}{c}{$s_{3}^{\prime}$} \\
	 \hline
	 $c_{1}$ &  3.1& -31.0 &  -1.5&  0.9& -7.4&  20.0&  5.7&  6.4&  -1.5&  0.9& -6.9\\
	 $c_{2}$  &    &  71.3& -21.2 &  2.1&  7.4&   -2.0 &  30.1&  24.9& 21.2&   2.1&  8.7\\
	 $c_{3}$   &    &     & -55.4 & 6.7& -7.0 & -13.0 &   18.8& 57.4&  -0.9 &  6.6& 5.6\\
	 $s_{1}$  &    &       &   &-30.4& -23.8&  -0.5 & -18.9 & -19.3 & 9.9&  -30.4& -23.6\\
	 $s_{2}$ &    &       &     &  &  29.2&  1.2& 1.1&  -0.4& -30.4& 90.0 & 31.6	\\
	 $s_{3}$ &    &       &     &    &   &   -8.0 & 5.9& -13.3 & 23.9&   20.0& 67.0\\
	 $c_{1}^{\prime}$ &    &       &     &    &     &    & 0.8& 5.7&  -0.5 & 1.2& -7.6\\
	 $c_{2}^{\prime}$ &    &       &     &    &     &      &    & 23.8& -18.9&    1.1& 7.0	\\
	 $c_{3}^{\prime}$ &    &       &     &    &     &      &      &   & -19.4&  -0.4& -13.3	\\
	 $s_{1}^{\prime}$ &    &       &     &    &     &      &      &     &   &  -30.4& -23.7	\\
	 $s_{2}^{\prime}$ &    &       &     &    &     &      &      &     &     &    & 31.6	\\
	 \hline	
	 \hline	 
\end{tabular}
\end{table*}

\begin{table*}[ht!]\scriptsize
	\centering
	\caption{Statistical correlation matrix (\%) for $K^{0}_{\mathrm{S}}K^{+}K^{-}$ equal-$\Delta\delta_{D}$ $\mathcal{N}=4$ binning.}	
	\label{tab:statcorrmat4bin}
	\setlength{\tabcolsep}{2.5pt}
	\begin{tabular}{l d d d d d d d d d d d d d d d d}
	\hline
	\hline
	 & \multicolumn{1}{c}{$c_{2}$} & \multicolumn{1}{c}{$c_{3}$} & \multicolumn{1}{c}{$c_{4}$} & \multicolumn{1}{c}{$s_{1}$} & \multicolumn{1}{c}{$s_{2}$} & \multicolumn{1}{c}{$s_{3}$} & \multicolumn{1}{c}{$s_{4}$} & \multicolumn{1}{c}{$c_{1}^{\prime}$} & \multicolumn{1}{c}{$c_{2}^{\prime}$} & \multicolumn{1}{c}{$c_{3}^{\prime}$} & \multicolumn{1}{c}{$c_{4}^{\prime}$} & \multicolumn{1}{c}{$s_{1}^{\prime}$} & \multicolumn{1}{c}{$s_{2}^{\prime}$} & \multicolumn{1}{c}{$s_{3}^{\prime}$} & \multicolumn{1}{c}{$s_{4}^{\prime}$}\\
	 \hline
	 $c_{1}$ &0.8&  2.3&  0.8& -2.3 &  0.2& -0.2 &  0.0&  98.6&  0.8&  2.3&  0.2&-2.3 &0.2&-0.2 &  0.0 \\
$c_{2}$ & &        0.3& -0.8 &  0.7& -3.0 &  0.0&  0.1&  0.8&  98.4&  0.3& -0.2 &  0.7&  -3.0 & 0.0&  0.1\\
$c_{3}$  &      &      & -0.1 & -0.6 &  0.2&  4.3&  0.0 &  2.4&  0.3&  97.0& -0.0 & -0.6 & 0.2&  4.3&  0.0\\
$c_{4}$    &      &      &      &  0.0& 0.1&  0.0 & -5.7 &  0.8& -0.8 & -0.1 & 24.9&  0.0& 0.1& 0.0& -3.6 \\
$s_{1}$    &      &      &      &      & -10.6 &  1.5& -0.1 & -2.4 &  0.7& -0.5 &  0.0&  99.7&  -10.6 &  1.5 & -0.1 \\
$s_{2}$    &      &      &      &      &      &  4.2& -2.6 &  0.2& -3.0 &  0.2&  0.0& -10.6 & 99.9&  4.2& -1.7 \\
$s_{3}$    &      &      &      &      &      &      &  0.0 & -0.2 &  0.0&  4.2&  0.0 &  1.5& 4.2& 99.8&  0.0 \\
$s_{4}$    &      &      &      &      &      &      &      & 0.0&  0.1&  0.0& -0.3 & -0.1 & -2.6 &  0.0& 66.6 \\
$c_{1}^{\prime}$    &      &      &      &      &      &      &      &      &  0.8&  2.3&  0.2& -2.4 &  0.2& -0.2 &  0.0 \\
$c_{2}^{\prime}$    &      &      &      &      &      &      &      &      &      &  0.3& -0.2 &  0.7&  -2.9 &  0.0&  0.1 \\
$c_{3}^{\prime}$    &      &      &      &      &      &      &      &      &      &      & 0.0& -0.5 &   0.2&  4.2&  0.0 \\
$c_{4}^{\prime}$   &      &      &      &      &      &      &      &      &      &      &      &  0.0&   0.0&  0.0&  0.8 \\  
$s_{1}^{\prime}$    &      &      &      &      &      &      &      &      &      &      &      &      &  -10.6 &  1.5& -0.1 \\
$s_{2}^{\prime}$    &      &      &      &      &      &      &      &      &      &      &      &      &       &  4.2& -1.7 \\
$s_{3}^{\prime}$    &      &      &      &      &      &      &      &      &      &      &      &      &       &      & 0.0 \\
	 \hline
	 \hline
\end{tabular}
\end{table*}

\begin{table*}[ht!]\scriptsize
	\centering
	\caption{Systematic correlation matrix for $K^{0}_{\mathrm{S}}K^{+}K^{-}$ equal-$\Delta\delta_{D}$ $\mathcal{N}=4$ binning.}	
	\label{tab:systcorrmat4bin}
	\setlength{\tabcolsep}{2.5pt}
	\begin{tabular}{l d d d d d d d d d d d d d d d}
	\hline
	\hline
	 & \multicolumn{1}{c}{$c_{2}$} & \multicolumn{1}{c}{$c_{3}$} & \multicolumn{1}{c}{$c_{4}$} & \multicolumn{1}{c}{$s_{1}$} & \multicolumn{1}{c}{$s_{2}$} & \multicolumn{1}{c}{$s_{3}$} & \multicolumn{1}{c}{$s_{4}$} & \multicolumn{1}{c}{$c_{1}^{\prime}$} & \multicolumn{1}{c}{$c_{2}^{\prime}$} & \multicolumn{1}{c}{$c_{3}^{\prime}$} & \multicolumn{1}{c}{$c_{4}^{\prime}$} & \multicolumn{1}{c}{$s_{1}^{\prime}$} & \multicolumn{1}{c}{$s_{2}^{\prime}$} & \multicolumn{1}{c}{$s_{3}^{\prime}$} & \multicolumn{1}{c}{$s_{4}^{\prime}$}\\
	 \hline
	 $c_{1}$ &  2.8& 2.7 & -2.7&  0.2& -3.8 & -1.9&	 2.0& 2.2&	 2.4&	2.7&	-1.3 &	0.2& -3.8 &	-2.1 &	 1.8\\
$c_{2}$ & &   53.2&	-23.7 &	12.6 & 2.9& -7.7&	 16.3& -30.8 & 64.0&  49.1& -7.2 & -12.6 & 2.9& -9.0 &  11.3	\\
$c_{3}$ &      &      &  -24.8 &	 -28.5 & -5.0 & -16.7 & 22.9& -33.3 & 50.3& 34.3& -10.5 & -22.7 & -5.0  & -17.1 &  20.4\\
$c_{4}$    &      &      &       &  -18.9 &  13.3& 0.1&	 -65.2&  31.6& -18.0 & -22.3 & 58.0& -18.9 & 13.3& 1.1& -35.0 \\
$s_{1}$    &      &      &       &       & -51.3 & 3.1& -14.9 &	 2.6&	 1.0& 27.5&	-9.5 &	 39.0&	-51.3 & 3.1&	-12.1\\
$s_{2}$    &      &      &       &       &      & 3.9&	 3.7& 0.9& 3.0&	-4.8 &	 5.9&	-43.0 &  99.0& 3.9&	 2.4\\
$s_{3}$    &      &      &       &       &      &     & -13.8 &	 1.0&	-8.9 &	 -16.5 &	-1.5 &	 3.2&	3.9 &	93.0&	-8.8 \\
$s_{4}$    &      &      &       &       &      &     &      & -0.5 & 2.0&	 6.1&	-19.0 &	-1.8 &	 3.8&	-14.3 &	 39.8 \\
$c_{1}^{\prime}$    &      &      &       &       &      &     &      &      & -24.2 &	-29.1 & 19.4&	2.7& -1.1&	1.7&	-14.6	\\
$c_{2}^{\prime}$   &      &      &       &       &      &     &      &      &      & 50.4&	-14.4&	 12.8&	2.8& -7.4&	 15.8	\\
$c_{3}^{\prime}$    &      &      &       &       &      &     &      &      &      &      & -13.2 &	-27.5 &	-4.9 &	 -16.1 &	 20.9\\
$c_{4}^{\prime}$   &      &      &       &       &      &     &      &      &      &      &      & -9.6 &	 6.2& -3.5 &	-43.8   	\\
$s_{1}^{\prime}$    &      &      &       &       &      &     &      &      &      &      &      &      & -51.2 &	3.2&	-12.1	\\
$s_{2}^{\prime}$    &      &      &       &       &      &     &      &      &      &      &      &      &      & 3.9&	2.3	\\
$s_{3}^{\prime}$    &      &      &       &       &      &     &      &      &      &      &      &      &      &     &  -8.4  \\
	 \hline	
	 \hline	 
\end{tabular}
\end{table*}

\section{Combination of BESIII and CLEO results}
\label{app:combination}
 The results presented in this paper are compatible with those reported by the CLEO Collaboration \cite{jim}. Therefore, it is advantageous to perform an average of the two sets of results to get the best possible estimates of $c_i^{(\prime)}$ and $s_i^{(\prime)}$. Following Ref. \cite{lei}, the average is calculated by adding a multi-dimensional constraint term to the log-likelihood expression given in Eq. (\ref{eq:kskkextract}). The constraint term is defined as
\begin{equation}
\chi^{2}_{\rm avg} = \left({\bf P} - {\bf P}^{\rm CLEO}\right)^{T}{\bf V}^{-1}\left({\bf P}-{\bf P}^{\rm CLEO}\right),
\end{equation}
where ${\bf P}$ is the vector of 4$\mathcal{N}$ parameters in the fit, ${\bf P}^{\rm CLEO}$ is the vector of corresponding values reported by the CLEO Collaboration and ${\bf V}$ is the $4\mathcal{N}~\times~4\mathcal{N}$ combined statistical and systematic covariance matrix related to the CLEO measurements. The values of $c^{(\prime)}_{i}$  and $s^{(\prime)}_{i}$ obtained from the fit including the $\chi^2_{\rm avg}$ term are given in Table ~\ref{tab:worldaverage}. The uncertainty returned by the fit only includes the BESIII statistical and total CLEO uncertainties. Hence, to take account of the systematic uncertainties related to the BESIII measurement, the uncertainties reported in Table ~\ref{tab:worldaverage} are the sum in quadrature of those returned by the fit and the systematic uncertainties reported in Tables~\ref{tab:systsummary3bin}, \ref{tab:systsummary2bin} and \ref{tab:systsummary4bin}. Calculating the uncertainty in this way assumes the systematic uncertainties related to the BESIII and CLEO measurements are uncorrelated; this is a valid assumption because the systematic uncertainties are dominated by those related to the strong-phase measurements of $D\to K^{0}_{\rm S}\pi^+\pi^-$ used in each analysis, which are independent measurements from the respective experiments. The correlation matrices reported in Tables~\ref{tab:combinedcorrmat2bin}-\ref{tab:combinedcorrmat4bin} are obtained by summing the covariance matrix from the fit with the systematic covariance matrix.

\begin{table*}[hb!]
	\centering
	\caption{Results for  $c_{i}^{(\prime)}$ and $s_{i}^{(\prime)}$ from averaging the results from BESIII and CLEO. The uncertainties on the values of parameters are the  statistical uncertainties obtained from fit added in quadrature with the systematic uncertainties. }
	\vspace{0.2cm}
	\setlength{\tabcolsep}{6.5pt}
	\begin{tabular}{c c c c c c c}
		\hline
		\hline
		$\mathcal{N}$ & $i$ & $c_{i}$ & $s_{i}$ & $c_{i}^{\prime}$ & $s_{i}^{\prime}$\\
		\hline
		2 & 1 & $\phantom{-}0.713 \pm 0.032$ & $\phantom{-}0.107 \pm 0.132$ & $\phantom{-}0.737 \pm 0.032$ & $\phantom{-}0.116 \pm 0.132$ \\
		  & 2 & $-0.758 \pm 0.037$ & $\phantom{-}0.394 \pm 0.173$ & $-0.782 \pm 0.033$ & $\phantom{-}0.473 \pm 0.174$\\
		  \hline
		3 & 1 & $\phantom{-}0.738 \pm 0.030$ & $\phantom{-}0.112 \pm 0.102$ & $\phantom{-}0.765 \pm 0.030$ & $\phantom{-}0.111 \pm 0.102$\\
		  & 2 & $-0.573 \pm 0.044$ & $\phantom{-}0.550 \pm 0.113$ & $-0.503 \pm 0.044$ & $\phantom{-}0.574 \pm 0.113$ \\
		  & 3 & $-0.129 \pm 0.155$ & $-0.619 \pm 0.317$ & $-0.412 \pm 0.138$ & $\phantom{-}0.089 \pm 0.327$\\
		 \hline
		4 & 1 & $\phantom{-}0.796 \pm 0.030$ & $-0.082 \pm 0.173$ & $\phantom{-}0.817 \pm 0.030$ & $-0.080 \pm 0.173$ \\
		  & 2 & $-0.018 \pm 0.099$ & $\phantom{-}0.393 \pm 0.262$ & $\phantom{-}0.105 \pm 0.098$ & $\phantom{-}0.375 \pm 0.261$\\
		  & 3 & $-0.691 \pm 0.048$ & $\phantom{-}0.551 \pm 0.200$ & $-0.657 \pm 0.048$ & $\phantom{-}0.601 \pm 0.200$ \\
		  & 4 & $\phantom{-}0.183 \pm 0.182$ & $-0.646 \pm 0.415$ & $-0.321 \pm 0.185$ & $\phantom{-}0.218 \pm 0.438$ \\    
		 \hline
		\hline
	\end{tabular}
	\label{tab:worldaverage}
\end{table*}



\begin{table*}[ht!]
	\centering
	\caption{Correlation matrix (\%) of the combined BESIII and CLEO results for the $K^{0}_{\mathrm{S}}K^{+}K^{-}$ equal-$\Delta\delta_{D}$ $\mathcal{N}=2$ binning.}	
	\label{tab:combinedcorrmat2bin}
	\setlength{\tabcolsep}{2.5pt}
	\begin{tabular}{l d d d d d d d d }
	\hline
	\hline
	 & \multicolumn{1}{c}{$c_{2}$} & \multicolumn{1}{c}{$s_{1}$} & \multicolumn{1}{c}{$s_{2}$} & \multicolumn{1}{c}{$c_{1}^{\prime}$} & \multicolumn{1}{c}{$c_{2}^{\prime}$} & \multicolumn{1}{c}{$s_{1}^{\prime}$} & \multicolumn{1}{c}{$s_{2}^{\prime}$} \\
	 \hline
	 $c_{1}$ &  7.1 & -4.3& -0.8 & 94.4 & 5.3 & -4.4 & -0.8 \\
	 $c_{2}$ & & -3.5 & -2.8 & 6.8 & 65.9 & -3.6 & -2.9 \\
	 $s_{1}$ & & & -4.1 & -4.1 & -2.5 & 99.2 & -4.0  \\
	 $s_{2}$ & & & & -0.8 & -3.5 & -4.0 & 96.8 \\
	 $c_{1}^{\prime}$ & & & & & 5.8 & -4.1 & -0.8 \\
	 $c_{2}^{\prime}$ & & & & & & -2.3 & -3.4\\
	 $s_{1}^{\prime}$& & & & & & &  -3.9\\
	 \hline	
	 \hline	 
\end{tabular}
\end{table*}

\begin{table*}[ht!]
	\centering
	\caption{Correlation matrix (\%) of the combined BESIII and CLEO results for the $K^{0}_{\mathrm{S}}K^{+}K^{-}$ equal-$\Delta\delta_{D}$ $\mathcal{N}=3$ binning.}
	\label{tab:combinedcorrmat3bin}	
		\setlength{\tabcolsep}{2.5pt}
		\begin{tabular}{l d d d d d d d d d d d d }
	\hline
	\hline
	    & \multicolumn{1}{c}{$c_{2}$} &
	    \multicolumn{1}{c}{$c_{3}$} &
        \multicolumn{1}{c}{$s_{1}$} & 
        \multicolumn{1}{c}{$s_{2}$} & 
        \multicolumn{1}{c}{$s_{3}$} &
        \multicolumn{1}{c}{$c_{1}^{\prime}$} & \multicolumn{1}{c}{$c_{2}^{\prime}$} &
    \multicolumn{1}{c}{$c_{3}^{\prime}$} &
    \multicolumn{1}{c}{$s_{1}^{\prime}$} & \multicolumn{1}{c}{$s_{2}^{\prime}$} &
    \multicolumn{1}{c}{$s_{3}^{\prime}$} \\
	 \hline
	 $c_{1}$ & 3.6 & 0.6  &  1.8  & -13.8 & 0.6 & 98.3 & 3.8 & 1.1  & 1.8 & -13.8 & 0.7\\
	 $c_{2}$ &    & 4.1 &  1.9  & 4.4 & 1.7  & 3.7  &  93.0 & 2.5  & 6.1 & 4.4  & 1.6\\
	 $c_{3}$ &    &    & -5.4 & 1.2 & 0.9 & 1.1  & 1.4 & 35.8 & 1.3 & 1.1  & 1.9\\
	 $s_{1}$ &    &      &  & -1.7 & -2.4  & 2.1 &  2.4 &   -1.6  &   79.0 & -1.7 & -2.3\\
	 $s_{2}$ &    &      &    &  & 7.5 & -13.2 & 4.1  & 0.6 & -1.7 & 98.4 & 7.4\\
	 $s_{3}$ &    &      &    &    &  & 0.8 & 1.7 & -1.1 & 6.1 & 6.1 & 89.6\\
	 $c_{1}^{\prime}$ &    &      &    &    &    &  & 3.9  & 1.3 & 2.1 & -13.1 & 0.8\\
	 $c_{2}^{\prime}$ &    &      &    &    &    &    &  &2.7 & 2.4 & 4.1 & 1.6\\
	 $c_{3}^{\prime}$ &    &      &    &    &    &    &    &  & -1.6 & -1.7 & -2.3\\
	 $s_{1}^{\prime}$ &    &      &    &    &    &    &    &    &  &-1.7 & -2.3 \\
	 $s_{2}^{\prime}$ &    &      &    &    &    &    &    &    &    &  & 7.5\\
     \hline
	 \hline	 
\end{tabular}
\end{table*} 

\begin{table*}[ht!]\scriptsize
	\centering
	\caption{Correlation matrix (\%) of the combined BESIII and CLEO results for the $K^{0}_{\mathrm{S}}K^{+}K^{-}$ equal-$\Delta\delta_{D}$ $\mathcal{N}=4$ binning.}	
	\label{tab:combinedcorrmat4bin}
	\setlength{\tabcolsep}{2.5pt}
	\begin{tabular}{l d d d d d d d d d d d d d d d d}
	\hline
	\hline
	 & \multicolumn{1}{c}{$c_{2}$} & \multicolumn{1}{c}{$c_{3}$} & \multicolumn{1}{c}{$c_{4}$} & \multicolumn{1}{c}{$s_{1}$} & \multicolumn{1}{c}{$s_{2}$} & \multicolumn{1}{c}{$s_{3}$} & \multicolumn{1}{c}{$s_{4}$} & \multicolumn{1}{c}{$c_{1}^{\prime}$} & \multicolumn{1}{c}{$c_{2}^{\prime}$} & \multicolumn{1}{c}{$c_{3}^{\prime}$} & \multicolumn{1}{c}{$c_{4}^{\prime}$} & \multicolumn{1}{c}{$s_{1}^{\prime}$} & \multicolumn{1}{c}{$s_{2}^{\prime}$} & \multicolumn{1}{c}{$s_{3}^{\prime}$} & \multicolumn{1}{c}{$s_{4}^{\prime}$}\\
	 \hline
	 $c_{1}$ &   3.1 & 3.2 & 2.5 & -1.3 & -12.8 &  0.7 & 3.9 &98.6 & 3.2 & 3.1 & -0.6 & -1.4 &  -12.9 & 0.7 & 5.6 \\
$c_{2}$ & & 3.7 & -2.4 & 5.0 & -1.7 & -0.3 & 0.2 & 2.6 & 97.3 & 3.5 & -0.2 & 3.7 & -1.7 & -0.4 & -0.8\\
$c_{3}$  &      &      & -0.6 & -2.9 & -3.1 & -0.4 & 2.7 & 2.4 & 3.6 & 95.1 & -0.4 & -2.5 & -3.1 & -0.4 & 2.7\\
$c_{4}$    &      &      &      & 3.3 & 2.6 & 1.0 & -2.6 & 3.2 & -2.2 & -0.5 & 41.0 & 3.3 & 2.6 & 1.0 & 0.9 \\
$s_{1}$    &      &      &      &      & -39.4 & 14.3 & 10.4 & -1.4 & 4.5 & 0.8 & -2.1 & 94.2 &   -39.4  & 14.3 & 13.4 \\
$s_{2}$    &      &      &      &      &      & -8.0 & -3.8 & -12.5 & -1.7 & -3.1 & 0.2 & -38.2 & 99.6 & -8.0 & -3.6 \\
$s_{3}$    &      &      &      &      &      &      &  0.7 & 0.7 & -0.3 & -0.5 &  -0.7 & 14.3 & -8.0      & 99.3 & 2.2 \\
$s_{4}$    &      &      &      &      &      &      &      &  3.6 & -1.0 & 0.8 & -0.4 & 12.1 & -3.8  & 0.6 & 77.6 \\
$c_{1}^{\prime}$    &      &      &      &      &      &      &      &      &  2.8 & 2.4 & -0.1 & -1.4 & -12.6 & 0.8 & 4.7\\
$c_{2}^{\prime}$    &      &      &      &      &      &      &      &      &      &  3.6 & -0.4 & 5.1 & -1.7 & -0.3 & -0.5 \\
$c_{3}^{\prime}$    &      &      &      &      &      &      &      &      &      &      & -0.4 &  -2.9 & -3.1 & -0.5 & 2.8 \\
$c_{4}^{\prime}$   &      &      &      &      &      &      &      &      &      &      &      &  -2.1  &  0.2 &  -0.8 & -3.1\\  
$s_{1}^{\prime}$    &      &      &      &      &      &      &      &      &      &      &      &      &  -39.5  &  14.3 & 13.3\\
$s_{2}^{\prime}$    &      &      &      &      &      &      &      &      &      &      &      &      &       &  -8.0 & -3.6 \\
$s_{3}^{\prime}$    &      &      &      &      &      &      &      &      &      &      &      &      &       &      &  2.3 \\
	 \hline
	 \hline
\end{tabular}
\end{table*}


\begin{thebibliography}{99}
\bibitem{ckmmatrix} N. Cabibbo, Phys. Rev. Lett. {\bf 10}, 531 (1963); M. Kobayashi and T. Maskawa, Prog. Theor. Phys. {\bf 49}, 652 (1973).
\bibitem{GLW} M. Gronau and D. Wyler, Phys. Lett. B {\bf 265}, 172 (1991); M. Gronau and D. London, Phys. Lett. B {\bf 253}, 483 (1991).
\bibitem{bib:brod} J.~Brod and J.~Zupan, JHEP {\bf 1401}, 051 (2014).
\bibitem{bib:buras} M.~Blanke and A.~J.~Buras, Eur.\ Phys.\ J.\ C {\bf 79}, 
159 (2019).
\bibitem{bib:lenz} A.~Lenz and G.~Tetlalmatzi-Xolocotzi, arXiv:1912.07621 [hep-ph].
\bibitem{ADS} D.~Atwood, I.~Dunietz and A.~Soni,
  Phys.\ Rev.\ Lett.\  {\bf 78}, 3257 (1997); D.~Atwood, I.~Dunietz and A.~Soni, Phys. Rev. D {\bf 63}, 036005 (2001).
\bibitem{bondarK0Spipi} A. Bondar, Proceedings of BINP special analysis meeting on Dalitz analysis, unpublished (2002).
\bibitem{giri} A. Giri, Y. Grossman, A. Soffer, and J. Zupan, Phys. Rev. D {\bf 68}, 054018 (2003).
\bibitem{bellegamma} H. Aihara {\it et al.} (Belle Collaboration), Phys. Rev. D {\bf 85}, 112014 (2012).
\bibitem{babargamma} P.~del Amo Sanchez \textit{et al.} (BaBar Collaboration),
Phys. Rev. Lett. \textbf{105}, 121801 (2010).

\bibitem{LHCbgamma} R.~Aaij {\it et al.} (LHCb Collaboration), J. High Energy Phys. {\bf 8}, 176 (2018).
\bibitem{resmi} P.K Resmi {\it et al.} (Belle Collaboration), J. High Energy Phys. {\bf 10}, 178 (2019). 
\bibitem{lei} M. Abilikim {\it et al.} (BESIII Collaboration) Phys. Rev. Lett. {\bf 124}, 241802 (2020); Phys. Rev. D {\bf 124}, 241802, (2020).
\bibitem{jim} J. Libby {\it et al.} (CLEO Collaboration), Phys. Rev. D {\bf 82}, 112006 (2010).
\bibitem{guy} C. Thomas and G. Wilkinson, J. High Energy Phys. {\bf 10} 185 (2012).
\bibitem{babarmodel} P. del Amo Sanchez {\it et al.} (BaBar Collaboration), Phys. Rev. D. {\bf 78}, 034023 (2008).
\bibitem{battaglieri} M. Battaglieri {\it et al.}, Acta Phys. Polon. B {\bf 46} 257 (2015). 
\bibitem{amo} P. del Amo Sanchez {\it et al.} (BaBar Collaboration), Phys. Rev. Lett. {\bf 105}, 121801 (2010).
\bibitem{poluektov} A.~Poluektov {\it et al.} (Belle Collaboration), Phys. Rev. D {\bf 81}, 112002 (2010).
\bibitem{LHCb-upgrade} R.~Aaij {\it et al.} (LHCb Collaboration),
 arXiv:1808.08865 [hep-ex].
\bibitem{BelleII}
E.~Kou {\it et al.} (Belle II Collaboration), PTEP {\bf 2019}, no. 12, 123C01 (2019)
\bibitem{bondar} A. Bondar and A. Poluektov, Eur. Phys. J. C {\bf 47}, 347 (2006); A. Bondar and A. Poluektov, Eur. Phys. J. C {\bf 55}, 51 (2008).
\bibitem{flatte} S. M. Flatté, Phys. Lett. B {\bf 63}, 224 (1976).
\bibitem{chris_thesis} C.~Thomas  (D.Phil. Thesis), University of Oxford (2011), (Retrieved from \url{http://cds.cern.ch/record/1427272}).
\bibitem{bepcluminosity} M. Ablikim {\it et al.} (BESIII Collaboration), Chin. Phys. C {\bf 37}, 123001 (2013).
\bibitem{bes3detector} M. Ablikim {\it et al.} (BESIII Collaboration), Nucl. Instrum. Methods Phys. Res., Sect. A {\bf 614}, 345 (2010).
\bibitem{geant4} S. Agostinelli {\it et al.} (GEANT4 Collaboration), Nucl. Instrum. Methods Phys. Res., Sec. A {\bf 506}, 250 (2003); J. Allison {\it et al.}, IEEE Trans. Nucl. Sci. {\bf 53}, 270 (2006); Z. Y. Deng {\it et al.}, High Energy Physics and Nuclear Physics {\bf 30}, 371 (2006).
\bibitem{kkmc} S. Jadach, B. F. L. Ward, and Z. Was, Comput. Phys. Commun. {\bf 130}, 260 (2000).
\bibitem{evtgen}D. J. Lange, Nucl. Instrum. Meth. A {\bf 462}, 152 (2001).
\bibitem{pdg} P.A. Zyla {\it et al.} (Particle Data Group), Prog. Theor. Exp. Phys. {\bf 2020}, 083C01 (2020).
\bibitem{lundcharm} J. C. Chen {\it et al.}, Phys. Rev. D {\bf 62}, 034003 (2000).
\bibitem{photos}E. Barberio and Z. Was, Comput. Phys. Commun. {\bf 79}, 291
(1994).
\bibitem{danthesis} D.~Ambrose (Ph.D. Thesis), University of Rochester (2014), (Retrieved from \url{https://inspirehep.net/literature/1669241}).

\bibitem{minakshi} S.~Malde {\it et al.}, Phys. Lett. B {\bf 747}, 9 (2015).

\bibitem{argus} H. Albrecht {\it et~ al.} (ARGUS Collaboration), Phys. Lett. B {\bf 241}, 278 (1990).
\bibitem{besNDDbar} M. Ablikim {\it et al.} (BESIII Collaboration), Chin. Phys. C {\bf 42}, 083001 (2018).
\bibitem{bigi} I.I. Bigi and H. Yamamoto, Phys. Lett. B {\bf 349}, 363 (1995).
\bibitem{cleo} D.~M.~Asner {\it et al.} (CLEO  Collaboration), Phys. Rev. D {\bf 78}, 012001 (2008).
\bibitem{briere} R. A. Briere (CLEO Collaboration), Phys. Rev. D. {\bf 80}, 032002 (2009).
\bibitem{coherencefactor} D. Atwood and A. Soni, Phys. Rev. D {\bf 68}, 033003 (2003).
\bibitem{hfag}Y. Amhis {\it et al.} (Heavy Flavor Averaging Group), Eur. Phys. J. C {\bf 77}, 895 (2017); and updates at \url{https://hflav.web.cern.ch}
\bibitem{libbycoherencepaper}J. Libby {\it et al.}, Phys. Lett. B {\bf 757}, 520 (2016); Phys. Lett. B {\bf 765}, 402 (2017).
\bibitem{topology} X.~Zhou, S.~Du, G.~Li and C.~Shen, arXiv:2001.04016 [hep-ex].
\bibitem{minuit} F. James and M. Roos, Comput. Phys. Commun. {\bf 10}, 343 (1975).
\bibitem{bes3whitepaper} M.~Ablikim {\it et al.} (BESIII Collaboration), Chin. Phys. C {\bf 44}, 040001 (2020).
\bibitem{lhcb} R. Aaij {\it et al.} (LHCb Collaboration), Phys. Rev. Lett. {\bf 122}, 231802 (2019).
\bibitem{bib:bes3k0skk} M.~Ablikim {\it et al.} (BESIII Collaboration), arXiv:2006.02800 [hep-ex].
\end{thebibliography}
\end{document}